\pdfoutput=1

\documentclass[ amsmath, amssymb, aps, prd, twocolumn, lengthcheck, sort&compress, nofootinbib, letterpaper ]{revtex4}

\usepackage{amsmath}
\usepackage{amssymb}
\usepackage{graphicx}
\usepackage{dsfont}
\usepackage{dcolumn}
\usepackage{units}
\usepackage{bm}
\usepackage{wasysym}
\usepackage{multirow}
\usepackage{slashed}
\usepackage[usenames]{color}
\usepackage[colorlinks=true,linkcolor=blue,citecolor=blue,urlcolor=blue]{hyperref}

      \newcommand{\vect}[1]{\bm{#1}}

      \newcommand{\mC}{\mathcal{C}}
      
      \newcommand{\mI}{\mathcal{I}}
      \newcommand{\mK}{\mathcal{K}}
      
      \newcommand{\mP}{\mathcal{P}}
      
      \newcommand{\mV}{\mathcal{V}}
      \newcommand{\mT}{\mathcal{T}}

      \newcommand{\be}{\begin{equation}}
      \newcommand{\ee}{\end{equation}}

      \DeclareMathOperator\atanh{atanh}

      \def\longlonglongrightarrow{
      \relbar\joinrel\relbar\joinrel\relbar\joinrel\relbar\joinrel\relbar\joinrel\relbar\joinrel\rightarrow}

      \definecolor{violet}{RGB}{111,0,255}
      \definecolor{webgreen}{rgb}{0,0.75,0}
      \definecolor{webred}{rgb}{0.75,0,0}
      \definecolor{webblue}{rgb}{0,0,0.75}
      \definecolor{darkblue}{rgb}{0,0,0.6}
      \definecolor{darkgreen}{rgb}{0,0.5,0.5}
      \definecolor{darkpurple}{rgb}{0.5,0,0.5}
      \definecolor{darkorange}{rgb}{1,0.5,0}
      \definecolor{darkgrey}{rgb}{0.4,0.4,0.4}
      \definecolor{lgray}{rgb}{0.95,0.95,0.95}
      \definecolor{lgreen}{rgb}{0.95,1.00,0.90}
      \definecolor{lred}{rgb}{1.00,0.90,0.80}
      \definecolor{lblue}{rgb}{0.2,0.35,1.00}
      \definecolor{shadecolor}{rgb}{1.00,0.92,0.82}

\begin{document}

         \title{Scattering amplitudes and contour deformations  }

       \author{Gernot~Eichmann$^1$}
       \author{Pedro~Duarte$^1$}
       \author{M.~T.~Pe\~na$^1$}
       \author{Alfred~Stadler$^{1,2}$}

\affiliation{$^1$CFTP and Departamento de Fis\'ica, Instituto Superior T\'ecnico, Universidade de Lisboa, 1049-001 Lisboa, Portugal  \\
             $^2$Departamento de Fis\'ica, Universidade de \'Evora, 7000-671 \'Evora, Portugal}

         \date{\today}

         \begin{abstract}
              We employ a scalar model to exemplify the use of contour deformations when
              solving Lorentz-invariant integral equations for scattering amplitudes.
              In particular, we calculate the onshell $2\to 2$ scattering amplitude
              for the scalar system.
              The integrals produce branch cuts in the complex plane of the integrand which prohibit
              a naive Euclidean integration path.
              By employing contour deformations, we can also access the kinematical regions
              associated with the scattering amplitude in Minkowski space.
              We show that in principle a homogeneous Bethe-Salpeter equation, together with
              analytic continuation methods such as the Resonances-via-Pad\'e method,
              is sufficient to determine the resonance pole locations on the second Riemann sheet.
              However, the scalar model investigated here does not produce resonance poles above threshold but instead virtual states
              on the real axis of the second sheet, which pose difficulties for analytic continuation methods.
              To address this, we calculate the scattering amplitude on the second sheet directly
              using the two-body unitarity relation which follows from the scattering equation.
         \end{abstract}


         \maketitle


    \renewcommand{\arraystretch}{1.0}

     \section{Introduction}

     The study of resonances is a central task in the nonperturbative treatment of quantum field theories.
     Among the observable states of a theory are bound states but also unstable resonances, which appear as bumps in cross sections
     and correspond to poles in the complex momentum plane on higher Riemann sheets.
     Among the prominent examples in QCD are the $\sigma$ meson, whose resonance pole position 
     is now well-established~\cite{Pelaez:2015qba},
     the excitations of light baryons,
     where multichannel partial-wave analyses of experimental precision data
     have led to the addition of several new states to the PDG~\cite{Tanabashi:2018oca},
     or the recently observed pentaquark states,
     where the nature of the neighboring peaks is still under debate~\cite{Aaij:2015tga,Aaij:2019vzc}.

     The theoretical investigation of scattering amplitudes and their resonance structure comes with technical challenges that
     appear in different guises. In a lattice formulation
     one calculates the finite-volume energy spectrum of the theory and extracts the resonance information
     through the Luescher method~\cite{Luscher:1986pf}, where
     current efforts focus on resonances above two- and three-particle thresholds;
     see e.g.~\cite{Hammer:2017kms,Briceno:2017max,Mai:2018djl,Hansen:2019nir,Jackura:2019bmu} and references therein.

     In continuum approaches, scattering amplitudes and their resonance information
     are accessible through scattering equations or Bethe-Salpeter equations (BSEs).
     Here the technical difficulties concern the numerical solution of four-dimensional scattering equations in the full kinematical domain
     and the extraction of resonance poles on higher Riemann sheets.
     On the one hand, the internal poles in the loop diagrams put restrictions on the accessible kinematic regions
     beyond which residue calculus or contour deformations into the complex momentum plane become inevitable.
     For example, without contour deformations one can only access low-lying excitation spectra or matrix elements in certain
     kinematic windows (the `Euclidean region').
     On the other hand, to extract the properties of resonances it is necessary to access unphysical Riemann sheets,
     whereas straightforward numerical calculations are restricted to the first sheet only.

     Both issues are typical obstacles in the functional approach of Dyson-Schwinger equations (DSEs) and BSEs,
     where one determines quark and gluon correlation functions and solves BSEs to arrive at hadronic observables.
     Progress has been made in the calculation of hadron spectra and matrix elements, see e.g.~\cite{Cloet:2013jya,Eichmann:2016yit,Sanchis-Alepuz:2017jjd,Burkert:2019bhp} and Refs. therein,
     but the treatment of resonances is still in its early stages.
     Beyond technical aspects there are also conceptual challenges:
     when constructing hadron matrix elements from quarks and gluons,
     the bound states (such as $\pi$ and $\rho$ mesons, $N$ and $\Delta$ baryons)
     and decay mechanisms that turn these bound states into resonances (e.g. $\rho\to\pi\pi$ or $\Delta\to N\pi$)
     must both emerge from the underlying quark-gluon interactions.
     The resonance mechanism corresponds to the dynamical emergence of internal hadronic poles
     in matrix elements ($\pi\pi$, $N\pi$, etc.), which are also responsible for the so-called `meson cloud' effects.
     Such properties have been studied by
     resumming quark and gluon topologies to intermediate meson propagators~\cite{Fischer:2008wy,Sanchis-Alepuz:2014wea,Williams:2018adr,Miramontes:2019mco}
     or by going to multiquark systems
     where these poles are generated dynamically~\cite{Heupel:2012ua,Eichmann:2015cra,Wallbott:2019dng}.
     In addition to internal hadronic poles, however, also the dynamical singularities encoded
     in the elementary quark and gluon correlation functions restrict the  kinematical domains of matrix elements
     and must be taken into account when calculating observables in the far spacelike, timelike or lightlike regions.

     In this work we focus on the methodological aspects, namely how to calculate scattering amplitudes
     in the kinematical domains where contour deformations become necessary (which is usually referred to as `going to Minkowski space'),
     and how that information can be used to extract the resonance information on higher Riemann sheets.
     Numerical contour deformations have been employed in the literature in the calculation of two- and three-point
     functions, see e.g.~Refs.~\cite{Maris:1995ns,Eichmann:2009zx,Strauss:2012dg,Windisch:2012zd,Windisch:2012sz,Pawlowski:2017gxj,Williams:2018adr,Miramontes:2019mco,Weil:2017knt};
     here we apply them to a four-point function in the form of a two-body scattering amplitude.

     \begin{figure}[t]
     \center{
     \includegraphics[width=0.93\columnwidth]{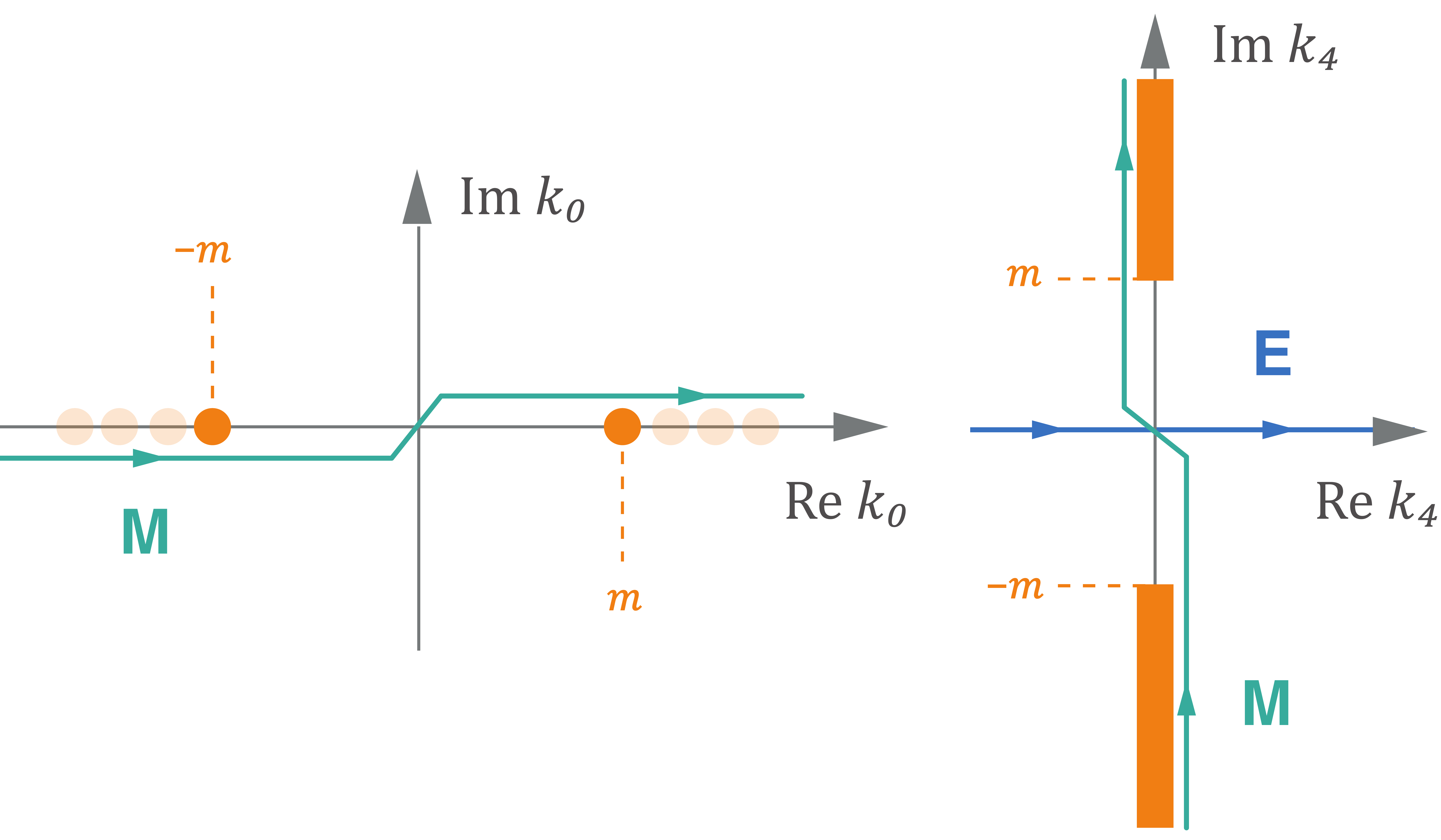}}
        \caption{Poles in the complex $k_0$ plane and cuts in the $k_4$ plane for a simple integral with one propagator pole.
                 M stands for the Minkowski path and E for the Euclidean contour.}
        \label{fig:1-pole}
     \end{figure}

     To illustrate the generic features, we exemplify the procedure for the simplest example, namely a scalar theory
     with a scalar exchange, which for a massless exchange particle becomes the well-known Wick-Cutkosky model~\cite{Wick:1954eu,Cutkosky:1954ru,Nakanishi:1969ph}.
     We calculate the $2\to 2$ scattering amplitude of the theory as well as its homogeneous Bethe-Salpeter (BS) amplitude
     and inhomogeneous BS vertex. We will see that in principle already the eigenvalues of the homogeneous BSE are sufficient to extract the resonance information.
     It turns out, however, that the model does not produce resonances above threshold but virtual states
     on the second Riemann sheet, which will pose difficulties for standard analytic continuation methods.
     Instead, we solve the scattering equation directly and employ the two-body unitarity property which
     allows us to calculate the scattering amplitude also on the second sheet.

     Recent progress has also been made
     in calculating propagators and BS amplitudes in Minkowski space, see~\cite{Carbonell:2010zw,Frederico:2013vga,Jia:2017niz,Leitao:2017esb,Leitao:2017qq,dePaula:2017ikc,Biernat:2018khd,Ydrefors:2019jvu,Frederico:2019noo,Solis:2019fzm} and references therein.
     Here we want to point out that there is no intrinsic difference between Euclidean and Minkowski space approaches:
     To obtain scattering amplitudes in the complex plane,
     contour deformations are necessary both in a Euclidean and Minkowski metric.
     When implemented properly, the resulting amplitude
     obtained with a Euclidean path deformation is identical to the Minkowski space amplitude.
     We discuss this in Sec.~\ref{sec:contour-deformations-intro} and use a Euclidean metric for the remainder of this work;
     Euclidean conventions can be found in Appendix A of Ref.~\cite{Eichmann:2018ytt}.

     The paper is organized as follows. After illustrating contour deformations for simple examples in Sec.~\ref{sec:contour-deformations-intro},
     we establish the scalar model that we employ in Sec.~\ref{sec:model}. In Sec.~\ref{sec:bse} we solve its homogeneous and inhomogeneous BSEs,
     explain the contour deformation procedure and the analytic continuation to the second Riemann sheet.
     In Sec.~\ref{sec:sc-eq} we solve the $2\to 2$ scattering equation of the theory, discuss the two-body unitary relation and present our results.
     We conclude in Sec.~\ref{sec:summary}. Technical details on contour deformations are relegated to Appendix~\ref{app:cd}.

     \begin{figure}[b]
     \center{
     \includegraphics[width=1\columnwidth]{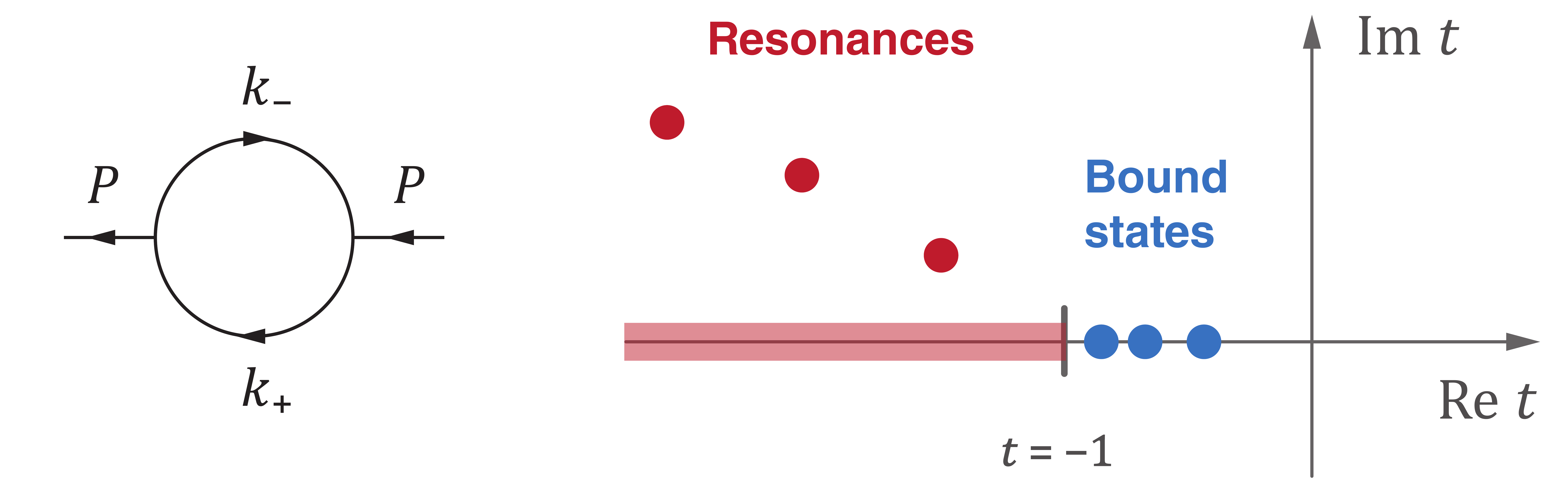}}
        \caption{Complex $t$ plane for a typical current correlator. The leading perturbative loop diagram
                 only produces the cut.}
        \label{fig:t-plane}
     \end{figure}

     \section{Contour deformations}\label{sec:contour-deformations-intro}

            To motivate the idea of contour deformations, we first illustrate
            the problem with simple examples.
            To begin with, consider an integral with only one propagator pole in the loop such as e.g. in
            a Fourier transform:
            \begin{equation}
                i\!\int d^4k\,\frac{1}{k^2-m^2+i\epsilon} \dots =
                i\!\int d^3k \! \int\limits_{-\infty}^\infty \!\! dk_0\,\frac{1}{k_0^2-\omega^2+i\epsilon} \dots
            \end{equation}
            where $\omega=\sqrt{\vect{k}^2+m^2}$.
            The integrand has poles on the real $k_0$ axis, whose locations depend on $\vect{k}^2$ and start at $k_0 = \pm m$
            (see Fig.~\ref{fig:1-pole}).
            In the standard Minkowski treatment one exploits the $i\epsilon$ term to shift the poles away from the real axis,
            performs the $k_0$ integration by closing the integration contour at complex infinity,
            picks up the appropriate residues, and finally integrates over the $\vect{k}^2$ dependence of those residues.

            In Euclidean conventions one defines $k_4 = ik_0$, but because real and momentum space rotate in opposite directions
            one has $d^4k_E = -i d^4k$.
            The integral then becomes
            \begin{equation}
               \int d^3k \! \int\limits_{-\infty}^\infty \!\! dk_4\,\frac{1}{k_4^2 + \omega^2} \dots =
               \int d^4k_E\,\frac{1}{k_E^2+m^2} \dots\,,
            \end{equation}
            where the integration proceeds from left to right in the complex $k_4$ plane
            and the poles lie on the imaginary axis.

            Now suppose we interchange the $d^3k$ and $dk_4$ integrations and integrate over $d^3k$ first.
            Along the former pole positions we now find branch cuts in the complex $k_4$ plane
            starting at $k_4 = \pm im$, illustrated in the right panel of Fig.~\ref{fig:1-pole}, and
            instead of closing the contour analytically we
            integrate numerically by avoiding the cuts. Clearly, the two strategies are equivalent: if the integration path
            crossed a cut, one would not pick up all the poles and obtain a wrong result.

            The Euclidean and Minkowski paths give identical results in this example. One can rotate the Minkowski path
            counterclockwise because there are no further poles in its way and the opposite integration direction
            is canceled by the sign in the integration measure --- a Wick rotation is possible.
            In the following we drop the subscript `E' and continue to work with Euclidean conventions.

            Now consider an integral with two poles in it, such as the loop diagram in Fig.~\ref{fig:t-plane}:
            \begin{equation}\label{int-2-poles}
              \mathcal{I}(t) = \int d^4k\,\frac{1}{k_+^2+m^2}\,\frac{1}{k_-^2+m^2}\,.
            \end{equation}
            The total momentum is $P^\mu$, the internal momenta are $k_\pm^\mu = k^\mu \pm P^\mu/2$,
            and we defined\footnote{Note that $t$ is related to the usual definition of the Mandelstam variable $\tilde s$ through
            $\tilde s=-4m^2 t$. We adapted our notation to the Compton scattering kinematics in Sec.~\ref{sec:half-offshell-amp};
            in the following we therefore refer to the resonant channel as the $t$ channel
            and to the crossed channels as $s$ and $u$ channels.}
            the dimensionless variable $t = P^2/(4m^2)$. The integral is Lorentz invariant
            and thus only depends on $t$. This is the simplest example of a two-point correlation function
            like a self-energy or vacuum polarization, which in principle can produce the singularity structure
            shown in Fig.~\ref{fig:t-plane}. Bound states appear on the negative real axis of $t$
            and resonances above the threshold $t<-1$ on higher Riemann sheets.
            The perturbative integral~\eqref{int-2-poles} can at best produce a two-particle cut but if the
            internal propagators and vertices were dressed and non-perturbative, the
            integral could also generate bound-state and resonance poles.

     \begin{figure}[t]
     \center{
     \includegraphics[width=0.5\columnwidth]{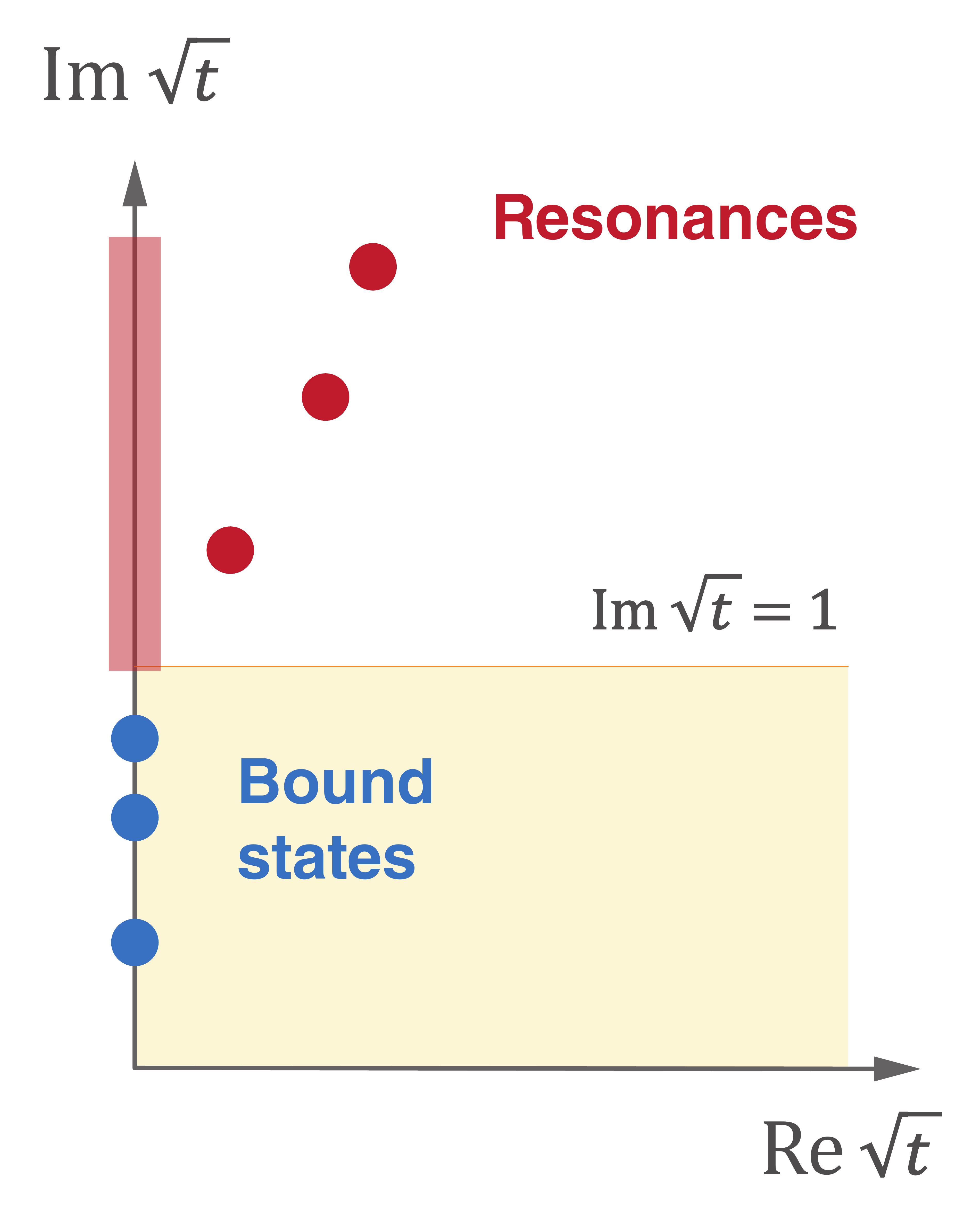}}
        \caption{Singularity structure in the complex $\sqrt{t}$ plane.}
        \label{fig:sqrt-t-plane}
     \end{figure}

            Because $\mathcal{I}(t)=\mathcal{I}^\ast(t^\ast)$ is an analytic function, it is sufficient to consider the upper half plane in $t$ only:
            The real part is symmetric around the real axis and the imaginary part is antisymmetric.
            It is then more convenient to plot the function in the complex $\sqrt{t}$ plane, which confines
            it to the upper right quadrant (Fig.~\ref{fig:sqrt-t-plane}). In this case the bound states
            appear on the imaginary axis below threshold ($\text{Im}\sqrt{t} < 1$), the cut starts
            at the threshold and the resonances lie above threshold on a higher Riemann sheet.
            In this way one can directly read off the real and imaginary parts of the masses $M_i$, which appear at
            $\text{Im}\sqrt{t} = \text{Re}\,M_i/(2m)$.

\begin{figure}[t]
     \center{
     \includegraphics[width=1\columnwidth]{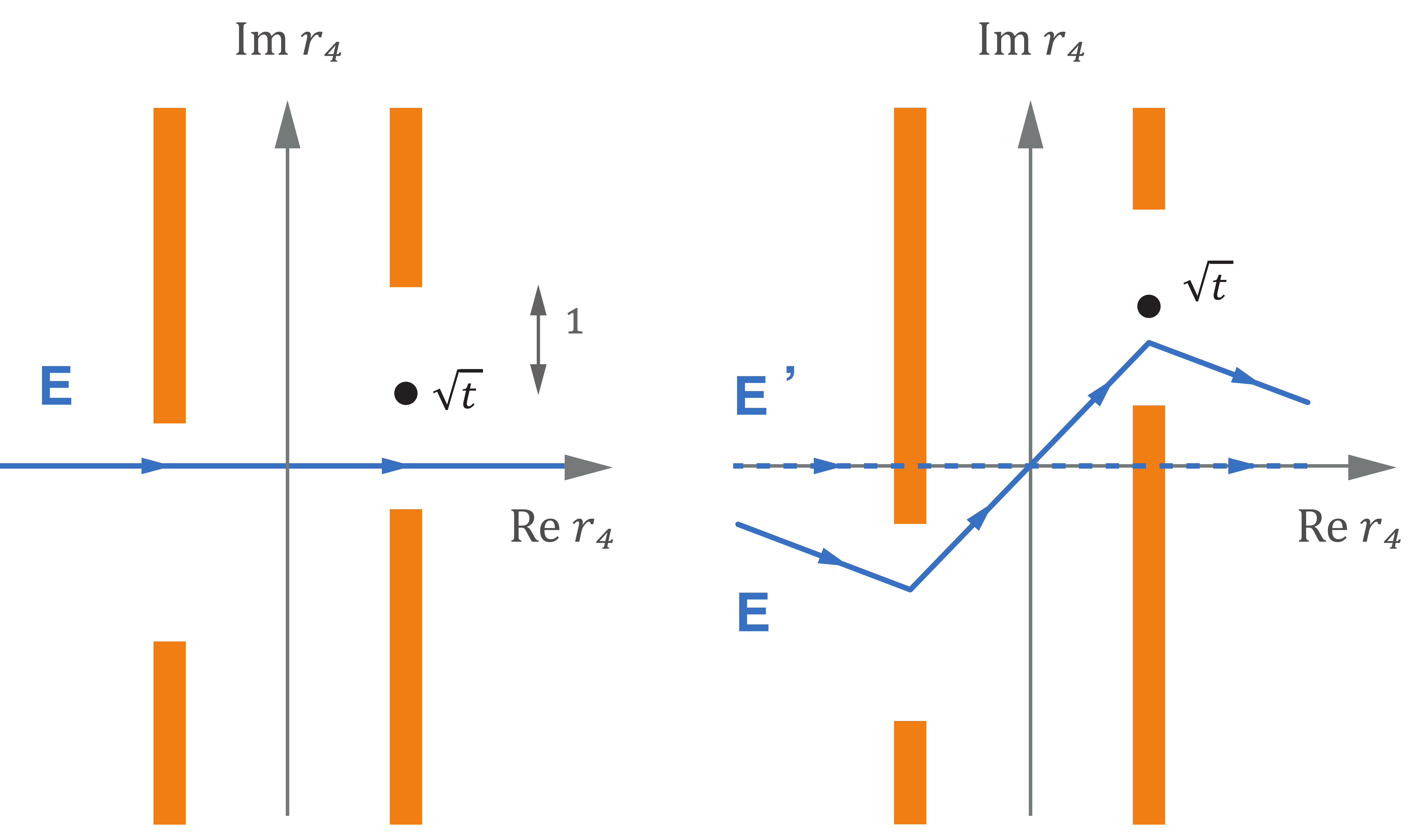}}
        \caption{Cuts in the complex $r_4$ plane for an integral with two poles.
                 In the left panel $\text{Im}\sqrt{t}<1$ and thus a straight integration path is sufficient;
                 in the right panel $\text{Im}\sqrt{t}>1$, which requires a contour deformation.}
        \label{fig:r4-plane-1}
     \end{figure}

            Suppose we want to calculate $\mathcal{I}(t)$ for some $t \in \mathds{C}$.
            Fig.~\ref{fig:r4-plane-1} shows the resulting cuts in the complex plane of $r_4 = k_4/m$.
            There are four vertical cuts centered around the external point $\pm \sqrt{t}$.
            Since we divided out the mass, the vertical distance between $\sqrt{t}$ and the onset of the cuts is equal to 1.
            As before, the Euclidean integration path proceeds from left to right.

            If $\text{Im}\sqrt{t} > 1$, however, the cuts cross the real axis
            and the straight Euclidean path (we denote it by E') would cross the cuts.
            Hence we must deform the contour to avoid the cuts: The correct Euclidean path is E.
            As a consequence, E = E' only below the threshold $\text{Im}\sqrt{t} < 1$,
            i.e., in the colored region in Fig.~\ref{fig:sqrt-t-plane},
            where a naive Euclidean integration is sufficient and gives the correct result.
            Above threshold, one has to deform the contour to obtain the correct value of the integral.
            The situation can be generalized to unequal masses or complex propagator poles,
             but the principle is the same: a straight Euclidean integration path is only valid in a limited domain of complex $t$.

\begin{figure}[t]
     \center{
     \includegraphics[width=1\columnwidth]{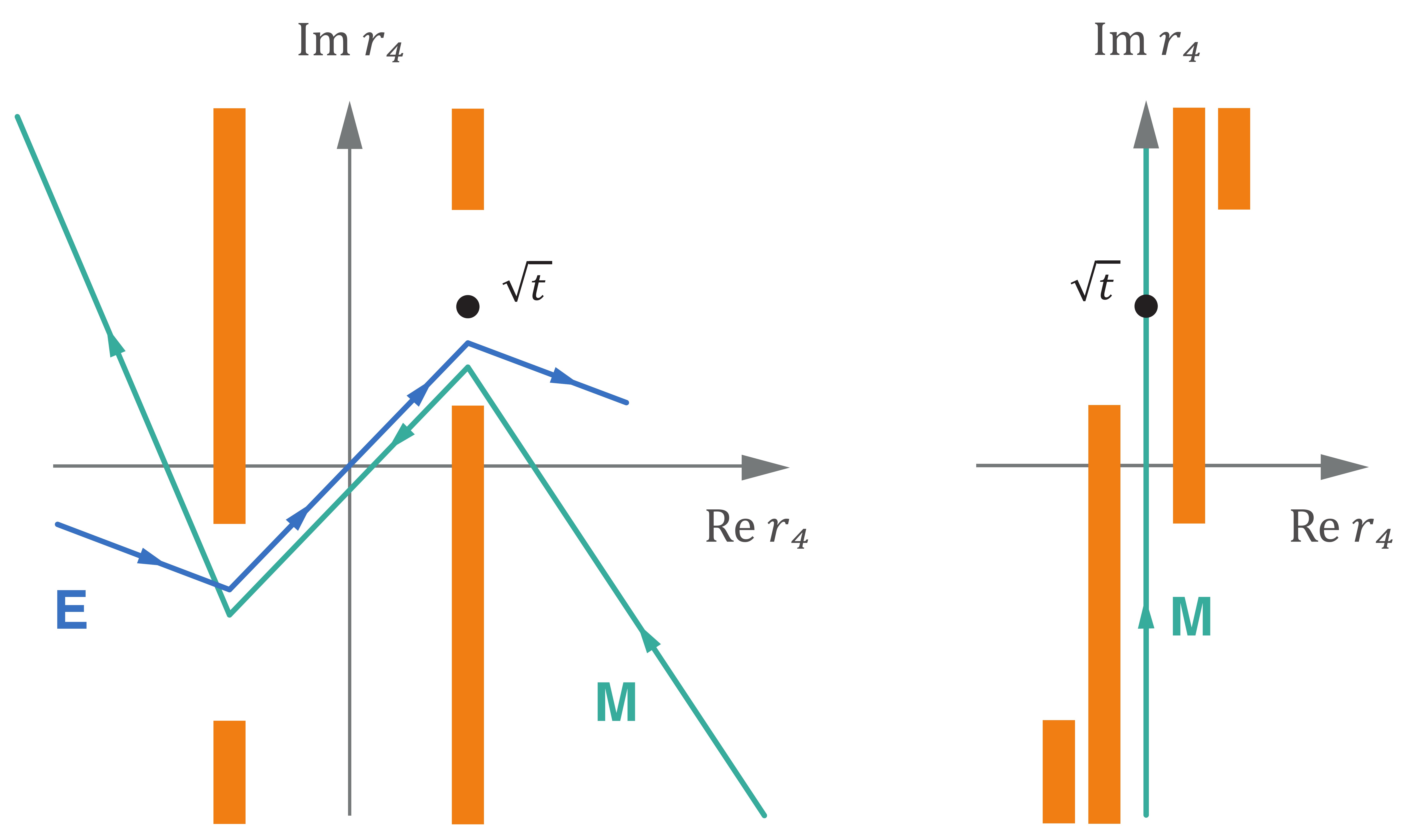}}
        \caption{\textit{Left:} Same as in Fig.~\ref{fig:r4-plane-1}; the Euclidean and Minkowski paths are equivalent.
                 \textit{Right:} Situation for $\text{Re}\sqrt{t}=0$, where the cuts are shifted by $i\epsilon$.}
        \label{fig:r4-plane-2}
     \end{figure}

            What would be the corresponding Minkowski path? Apparently it cannot proceed along the vertical axis
            as in Fig.~\ref{fig:1-pole}:
            It does not matter whether we start slightly on the right and end up slightly on the left
            because there are no singularities on the imaginary axis.
            In fact, the $i\epsilon$ prescription entails
            \begin{equation}
                \int\limits_{-\infty(1+i\epsilon)}^{\infty(1+i\epsilon)} \!\!\!\!\!\! dk_0 \quad \Leftrightarrow \quad
                 \int\limits_{-\infty(i-\epsilon)}^{\infty(i-\epsilon)} \!\!\!\!\!\! dr_4\,,
            \end{equation}
            since it originates from the need to isolate the interacting vacuum $|\Omega\rangle$ in a correlation function,
             \begin{equation*}
                 \sum_{n=0}^\infty e^{-iE_n T} |n\rangle\langle n|\Omega\rangle
                 \stackrel{T\to \infty(1-i\epsilon)}{\longlonglongrightarrow}
                 e^{-iE_0 T} |0\rangle\langle 0|\Omega\rangle\,,
             \end{equation*}
             and thereby remove the higher energy contributions $E_n$ of the free $n-$particle states $|n\rangle$.
             The integration path between $T=\pm \infty(1-i\epsilon)$ in the action of the quantum field theory
             thus corresponds to $k_0\to \pm \infty(1+i\epsilon)$ and $r_4 \to \pm\infty\,(i - \epsilon)$.
             Therefore, the proper Minkowski path is
             the diagonal line from bottom right to top left, which must \textit{also} be deformed to avoid the cuts,
             cf.~Fig.~\ref{fig:r4-plane-2}.

             That this is indeed the correct path can also be seen by putting the point $\sqrt{t}$ back onto the imaginary axis
             (right panel in Fig.~\ref{fig:r4-plane-2}).
             In that case all cuts also lie on the imaginary axis and one can use the $i\epsilon$ term to displace the cuts,
             while the Minkowski path is the straight line from bottom to top.
             As a consequence, all cuts in the upper half plane of $r_4$ are shifted to the right of the path and all cuts
             in the lower half plane to its left; closing the contour on either side gives the same result.
             This is also what happens in the left panel of Fig.~\ref{fig:r4-plane-2},
             where the upper cuts appear on the right of the Minkowski path and the lower cuts on its left ---
             but that path is just the same as the Euclidean contour.

             In general there is no intrinsic difference between quantities in `Euclidean' or `Minkowski space'.
             In both cases one has to deform integration contours to avoid cuts and the final result is the same.
             A naive Euclidean integration path would give the wrong result above threshold,
             whereas a naive Minkowski integration (in the sense of a straight vertical path)
             becomes meaningless once $t$ is complex. The $i\epsilon$ prescription in the action
             only tells us where the integration starts and ends; that one additionally has to deform contours
             at the level of correlation functions is implicit in their definition.
             Therefore, what we mean by `Euclidean conventions' is merely a Euclidean \textit{metric}
             with metric tensor $\delta^{\mu\nu}$ as opposed to a Minkowski metric $g^{\mu\nu}$.
             A collection of Euclidean conventions can be found e.g. in Appendix A of Ref~\cite{Eichmann:2018ytt}.

             As an aside, there is also no inherent problem with perturbation theory which is usually
             done in Euclidean space. The `Wick rotation' only amounts to rewriting the integrals in a Euclidean metric,
             which is always possible as long as the singularities of their integrands are kept in mind
             and there is no contribution from the contour at complex infinity.
             If one employs Feynman parametrizations, then the integrals in momentum space become integrals over Feynman parameters
             and one has to analyze the singularity structure of the integrands in Feynman parameter space instead of momentum space.

             \pagebreak

             In the remainder of this paper we will need contour deformations for integrals with three and four internal poles,
             which are also not just integrals but appear in integral equations.
             Because all expressions are Lorentz invariant, it is convenient
             to preserve manifest Lorentz invariance by splitting the loop integral into
             $k^2$ and a three-dimensional solid angle via hyperspherical variables:
             \begin{equation}\label{int-measure}
                 \int d^4k = \frac{1}{2}\int\limits_0^\infty dk^2\,k^2 \int\limits_{-1}^1 dz \sqrt{1-z^2}\int\limits_{-1}^1  dy\int\limits_0^{2\pi} d\psi\,.
             \end{equation}
             Equivalently, one could use the radial variable $k = \sqrt{k^2}$ such that $dk^2\,k^2/2 = dk\,k^3$.
             As a consequence, the cuts are no longer vertical lines in the complex $k^2$ plane but pick up more complicated shapes
             which we discuss in Sec.~\ref{sec:cd}. In any case, the strategy is the same: Depending on the external kinematics,
             one must deform the straight Euclidean integration path from $k^2=0$ to $k^2\to\infty$ to avoid the cuts in the complex $k^2$ plane.

     \begin{figure}[t]
     \center{
     \includegraphics[width=1\columnwidth]{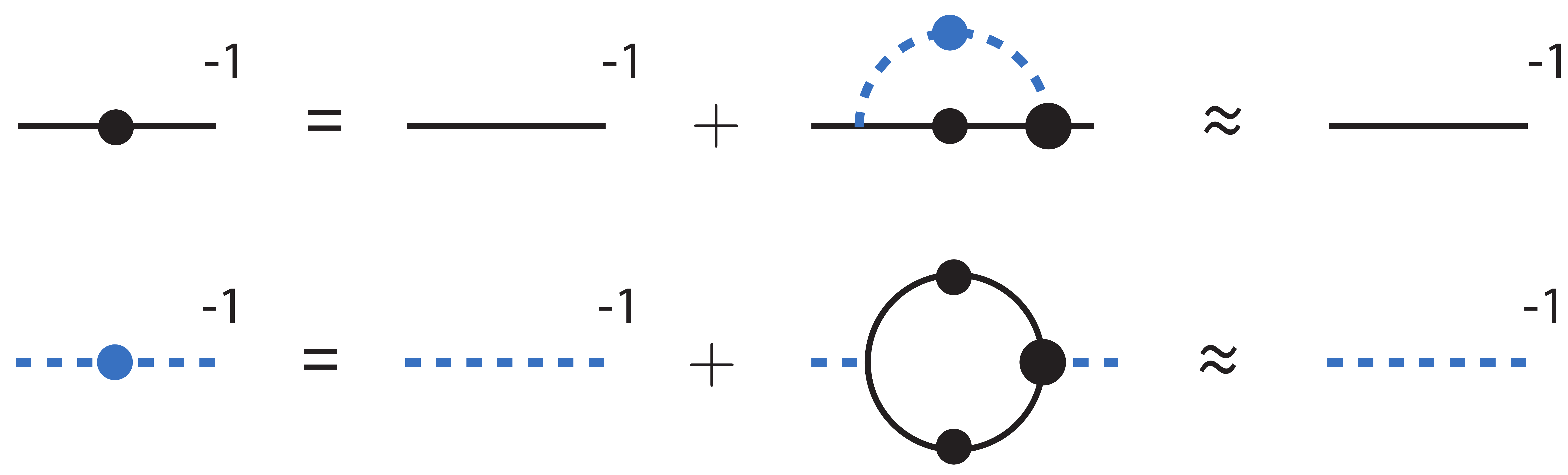}}
        \caption{Coupled propagator DSEs for the scalar model.}
        \label{fig:dse}
     \end{figure}

     \section{Scalar model} \label{sec:model}

             We consider the simplest scalar model that is capable of producing resonances:
             Two scalar particles $\phi$ and $\varphi$ with masses $m$ and $\mu$, respectively, and
             a three-point interaction $\sim g \phi\phi\varphi$. This leaves two parameters:
             \begin{equation}
                c = \frac{g^2}{(4\pi m)^2}\,, \qquad
                \beta = \frac{\mu}{m}\,.
             \end{equation}
             $\beta$ is the mass ratio and the coupling $g$ is dimensionful, so we defined a dimensionless coupling constant $c$.
             As a consequence, the mass $m$ drops out from all equations and only sets the scale.

\begin{figure}[t]
     \center{
     \includegraphics[width=1\columnwidth]{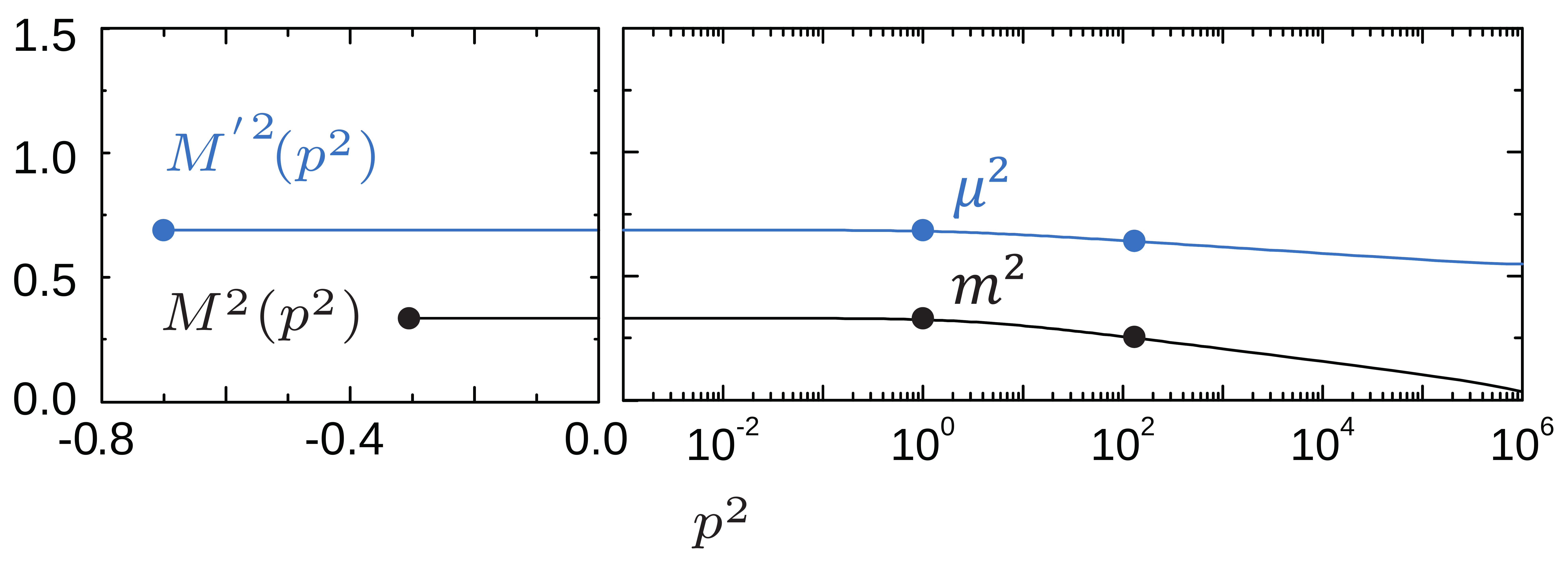}}
        \caption{Mass functions of the scalar theory in the timelike and spacelike region for three renormalization points
                 and fixed coupling $c$ (in arbitrary units).}
        \label{fig:mf}
     \end{figure}

             In principle one should dress the propagators by solving their coupled Dyson-Schwinger equations (DSEs)
             shown in Fig.~\ref{fig:dse}. However, tree-level propagators are good enough for our purposes
             because the dressing effects in the scalar theory are relatively small.
             If we define a mass function for each propagator,
             $M(p^2)$ and $M'(p^2)$, via
             \begin{equation}
             \begin{split}
                D(p^2) &= \frac{1}{Z}\,\frac{1}{p^2+M(p^2)^2}\,, \\
                D'(p^2)&= \frac{1}{Z'}\,\frac{1}{p^2+M'(p^2)^2}
             \end{split}
             \end{equation}
             where $Z$ and $Z'$ are the respective renormalization constants,
             then an exemplary solution of the coupled DSEs for a given value of $c$ (and using tree-level vertices only)
             is shown in Fig.~\ref{fig:mf}.
             The three dots on each curve correspond to three different renormalization points
             at which the renormalized masses $m$ and $\mu$ were specified as input; the renormalization constants $Z$ and $Z'$
             are determined in the solution process.
             One can see that both mass functions are essentially flat over
             a large momentum range and only begin to drop in the ultraviolet.

             If the coupling $c$ becomes large enough,
             both curves for the squared mass functions eventually cross zero and become negative
             in the ultraviolet. This reflects the vacuum instability of the $\phi\phi\varphi$ theory
             and implies that the model is only physically acceptable for small couplings,
             which  has also consequences for the so-called anomalous BSE solutions~\cite{Ahlig:1998qf}.
             Since we employ the scalar theory only as a toy model for calculating resonance properties,
             we will not further consider this point and restrict ourselves to tree-level propagators
             in what follows.

     \begin{figure}[b]
     \center{
     \includegraphics[width=0.97\columnwidth]{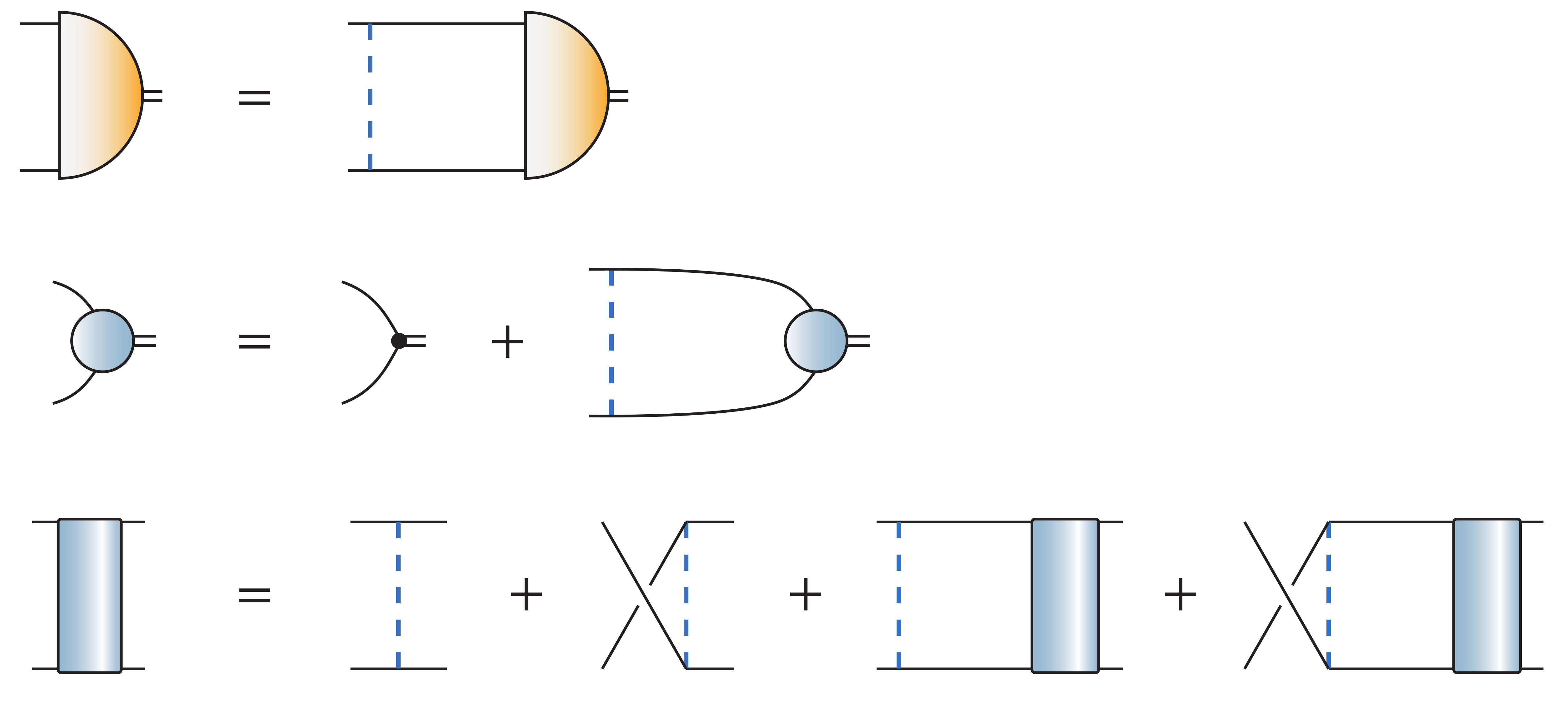}}
        \caption{From top to bottom: homogenous BSE, inhomogeneous BSE and scattering equation.}
        \label{fig:eqs}
     \end{figure}

             To extract the bound state and resonance properties of the model,
             we consider the three equations depicted in Fig.~\ref{fig:eqs}.
             The first is the homogeneous BSE for the BS amplitude $\Psi$,
             which in a compact form reads
             \begin{equation}
                \Psi = KG_0 \Psi\,.
             \end{equation}
             $K$ stands for the BS kernel and $G_0$ for the product of the two
             propagators. We employ a simple ladder exchange, which
             for a vanishing exchange-particle mass $\mu=0$
             becomes the Wick-Cutkosky model~\cite{Wick:1954eu,Cutkosky:1954ru,Nakanishi:1969ph}.
             The homogeneous BSE yields the eigenvalues of $KG_0$ for given quantum numbers $J^{PC}$,
             which depend on the total momentum variable $t \in \mathds{C}$.
             If an eigenvalue satisfies $\lambda_i(t)=1$, this corresponds to a pole
             in the scattering amplitude. Hence, in principle one can extract both the
             bound-state and resonance information from the homogeneous BSE.

             The second equation in Fig.~\ref{fig:eqs} is the inhomogeneous BSE for the vertex $\Gamma$:
             \begin{equation}\label{ibse-0}
                \Gamma = \Gamma_0 + KG_0 \Gamma\,.
             \end{equation}
             The three-point vertex is essentially the scattering amplitude in a given $J^{PC}$ channel:
             If $G$ denotes the full four-point function satisfying $G=G_0 + G_0 K G$, then
             the vertex is defined as $G_0 \Gamma = G\,\Gamma_0$, i.e., modulo external propagators
             it is the contraction of $G$ with the tree-level vertex $\Gamma_0$ carrying quantum numbers $J^{PC}$.
             Thus, it contains the bound-state and resonance poles in that channel.
             In a symbolic form one can write the solution of Eq.~\eqref{ibse-0} as
             \begin{equation}
                \Gamma = \frac{\Gamma_0}{1-KG_0}\,,
             \end{equation}
             which shows explicitly that whenever an eigenvalue of $KG_0$ becomes 1, one has found a pole
             in the vertex.

             Finally, the third equation in Fig.~\ref{fig:eqs} is the scattering equation for the full scattering amplitude $T$:
             \begin{equation}\label{sc-eq-0}
                T = K + KG_0 T\,,
             \end{equation}
             where $T$ is the connected and amputated part of $G$ defined via $G = G_0 + G_0 T G_0$.
             Since we denoted the total momentum variable by $t$, we will refer to the `horizontal' channel
             as the $t$ channel in the following. To ensure crossing symmetry in the $s$ and $u$ channels,
             we have symmetrized the kernel in the scattering equation; for the two preceding equations
             this is not necessary because both terms yield the same result. Symbolically, the solution
             of the scattering equation has the form
             \begin{equation}
                T = \frac{K}{1-KG_0}
             \end{equation}
             and therefore $T$ contains all singularities, including the exchange-particle poles
             encoded in the kernel $K$.

             In the following we investigate these equations in detail:
             the homogeneous and inhomogeneous BSEs in Sec.~\ref{sec:bse} and the scattering equation in Sec.~\ref{sec:sc-eq}.

     \section{Resonances from the BSE} \label{sec:bse}

     \subsection{Explicit form of the BSE}

             The homogeneous BSE reads explicitly:
             \begin{equation}\label{bse-hom}
                \Psi(q,P) = \int\limits_k  K(q,k)\,G_0(k,P)\,\Psi(k,P)\,,
             \end{equation}
             where $\int_k = d^4k/(2\pi)^4$ is the integral measure, $q$ is the relative and $P$ the total momentum,
             $k$ is the relative momentum in the loop, and the propagator momenta are $q_\pm = q \pm P/2$ and $k_\pm = k \pm P/2$.
             We restrict ourselves to scalar bound states with quantum numbers $J^{PC}=0^{++}$; in that case
             $\Psi(q,P)$ is a scalar as well.

             The internal propagators are given by
             \begin{equation}
               D(k_\pm) = \frac{1}{k_\pm^2+m^2} = \frac{1}{k^2+\tfrac{1}{4}P^2 + m^2 \pm k\cdot P}\,,
             \end{equation}
             so their product $G_0(k,P) = D(k_+)\,D(k_-)$ becomes
             \begin{equation}\label{G0-0}
               G_0(k,P) = \frac{1}{(k^2+\tfrac{1}{4}P^2+m^2)^2 - (k\cdot P)^2}\,.
             \end{equation}
             Symmetrizing the kernel is not necessary
             at this stage but we do it nevertheless for later convenience.
             The kernel is then the sum of $s$ and $u$-channel exchanges:
             \begin{equation}\label{kernel-0}
             \begin{split}
                 K(q,k) &= \frac{g^2}{2}\left[ \frac{1}{(q-k)^2+\mu^2} + \frac{1}{(q+k)^2 + \mu^2}\right] \\
                        &= \frac{g^2\,(q^2+k^2+\mu^2)}{(q^2+k^2+\mu^2)^2 - 4(q\cdot k)^2}\,.
             \end{split}
             \end{equation}

             Using the hyperspherical variables from Eq.~\eqref{int-measure}, we choose the rest frame
             \begin{equation}\label{frame-hyperspherical-0}
             \begin{split}
                P &= 2m\sqrt{t} \left[ \begin{array}{c} 0 \\ 0 \\ 0 \\ 1 \end{array}\right], \quad
                q = m\sqrt{X} \left[ \begin{array}{c} 0 \\ 0 \\ \sqrt{1-Z^2}  \\ Z \end{array}\right], \\
                k &= m\sqrt{x} \left[ \begin{array}{l} \sqrt{1-z^2}\,\sqrt{1-y^2}\,\sin\psi \\ \sqrt{1-z^2}\,\sqrt{1-y^2}\,\cos\psi \\ \sqrt{1-z^2} \,y \\ z \end{array}\right]
             \end{split}
             \end{equation}
             so that the radial integration variable is $k^2 = m^2 x$ and its external counterpart is $q^2 = m^2 X$.
             In principle, however, we never need to specify a frame if we define those variables in a Lorentz-invariant way:
             \begin{equation} \label{LI-variables-0}\renewcommand{\arraystretch}{1.4}
             \begin{array}{rl}
                P^2 &= 4m^2 \,t\,, \\
                q^2 &= m^2 X\,, \\
                k^2 &= m^2 x\,,
             \end{array}\quad
             \begin{array}{rl}
                q\cdot P &= 2m^2\sqrt{Xt}\,Z\,, \\
                k\cdot P &= 2m^2\sqrt{xt}\,z\,, \\
                q\cdot k &= m^2\sqrt{Xx}\,\Omega
             \end{array}
             \end{equation}
             where $\Omega = zZ + \sqrt{1-z^2}\sqrt{1-Z^2}\,y$\,.
             As a result, the kernel and two-body propagator become
            \begin{equation}\label{KG-2}
            \begin{split}
               K(X,x,\Omega) &= \frac{g^2}{m^2}\,\frac{X+x+\beta^2}{(X+x+\beta^2)^2 - 4Xx\,\Omega^2}\,, \\
               G_0(x,z,t) &= \frac{1}{m^4}\,\frac{1}{(x+t+1)^2-4xt\,z^2}
            \end{split}
            \end{equation}
            and the BSE takes the form
            \begin{equation}\label{bse-3}
            \begin{split}
               &\Psi(X,Z,t) = \frac{m^4}{(2\pi)^3}\,\frac{1}{2} \int\limits_0^\infty dx\,x\int \limits_{-1}^1 dz\sqrt{1-z^2} \\[-3mm]
                            &  \quad \times \int\limits_{-1}^1 dy \,K(X,x,\Omega)\,G_0(x,z,t)\,\Psi(x,z,t)
            \end{split}
            \end{equation}
            where the mass $m$ drops out.
            The integration over the angle $\psi$ is trivial because no Lorentz invariant depends on it and
            the integral over $y$ can be performed analytically, 
            \begin{equation}
                \int\limits_{-1}^1 dy\,\frac{1}{1-(a+b\,y)^2}   = \frac{1}{b}\left( \atanh \delta_+ + \atanh \delta_- \right),
           \end{equation}
           with $\atanh\delta = [\ln(1+\delta)-\ln(1-\delta)]/2$ and $\delta_\pm = b/(1\pm a)$.

             The inhomogeneous BSE has the same form as Eq.~\eqref{bse-3}
            if we replace $\Psi(X,Z,t) \to \Gamma(X,Z,t)$ and add an inhomogeneous term $\Gamma_0 = 1$
            on the right-hand side corresponding to quantum numbers $J^{PC} = 0^{++}$.

     \subsection{Contour deformation}   \label{sec:cd}

            To proceed, we analyze the singularity structure of the integrand in Eq.~\eqref{bse-3}.
            The equation is solved at fixed $t$, which remains an external parameter.
            After performing the $y$ and $z$ integrations, the poles in $K$ and $G_0$ produce
            cuts in the complex plane of the radial variable $x$,
             and if those cuts cross the positive real axis the `naive' integration path $0 < x < \infty$ is forbidden.
             Since we discuss the complex $\sqrt{t}$ plane instead of $t$, we also
             analyze the cut structure in the complex $\sqrt{x}$ plane; and
             we do not change the integration domain of the variable $z$ which remains in the interval $-1 < z < 1$.

            The singularities encoded in the quantities~\eqref{KG-2} can be written in a closed form by defining the function
            \begin{equation}\label{f-cut-0}
               f_\pm(t,\alpha,z) = \sqrt{t}\,z \pm i\sqrt{t(1-z^2)+\alpha^2}\,,
            \end{equation}
            where $t\in\mathds{C}$ and $\alpha>0$ are parameters and the resulting cut
            is parametrized by $-1 < z < 1$:
            \begin{itemize}
            \item[(1)] The propagator poles encoded in $G_0$ generate a cut $\sqrt{x}=\mathcal{C}_1 = f_\pm(t,1,z)$.
            \item[(2)] The exchange particle poles in $K$ produce a cut $\sqrt{x}=\mathcal{C}_3 =f_\pm(X,\beta,\Omega)$,
                       with $\Omega$ defined below~\eqref{LI-variables-0}.
            \item[(3)] We may need the vertex $\Gamma(X,Z,t)$ at the kinematical point where also the
                       constituent particles are onshell. In that case
                       \begin{equation}\label{onshell-0}
                          q_\pm^2 = -m^2 \;\; \Rightarrow \;\; Z=0, \; X=-(1+t)
                       \end{equation}
                       so that the kernel produces another cut at $\sqrt{x}=\mathcal{C}_2 =f_\pm(-(1+t),\beta,\sqrt{1-z^2}\,y)$.
                       The combination of $\mC_1$ and $\mC_2$ will become especially relevant for the onshell scattering amplitude in Sec.~\ref{sec:sc-eq}.
            \item[(4)] By means of the inhomogeneous BSE, the vertex $\Gamma(X,Z,t)$ develops bound-state poles and multi-particle cuts which are
                       functions of $t$ only. Since the equation is solved at a given $t$,
                       the only constraints that these singularities provide is that one
                       cannot obtain a solution directly along the cut
                       or at a pole location.
                       In practice we avoid the imaginary $\sqrt{t}$ axis and solve the equation
                       for $\text{Re}\sqrt{t}>0$, which circumvents both problems.
            \item[(5)] In principle the BS amplitude or vertex can dynamically develop singularities
                       in the relative momentum variables $X$ and $Z$. The variable $Z$ is protected
                       because we keep it in the interval $-1 < Z < 1$, and as long as the singularities
                       in $X$ only appear on the timelike $X$ axis they do not pose any restriction.
            \end{itemize}

     \begin{figure}[t]
     \center{
     \includegraphics[width=0.9\columnwidth]{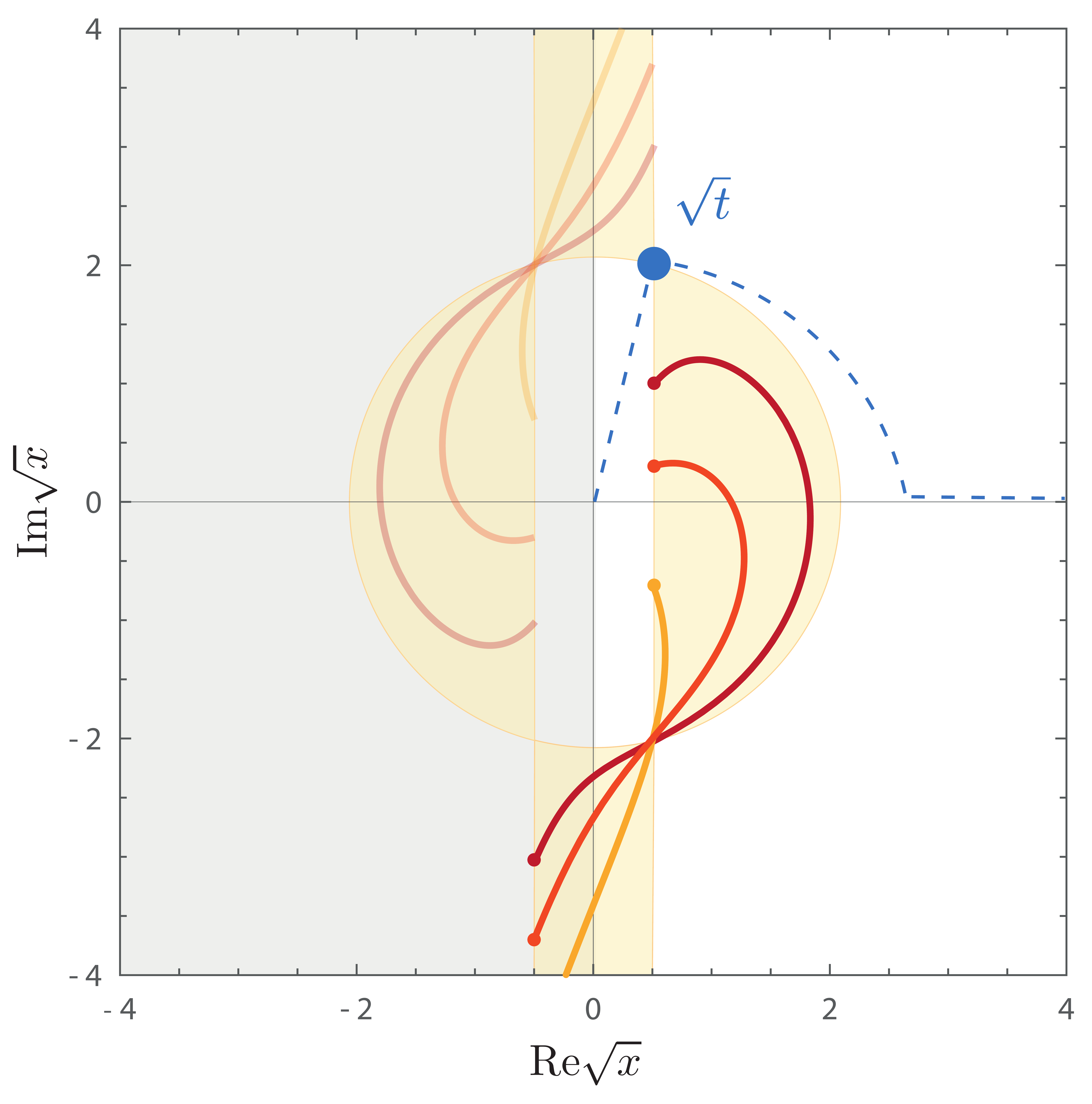}}
        \caption{The functions $f_\pm(t,\alpha,z)$ for $\sqrt{t}= 0.5+2i$ and three values $\alpha=1$, $1.7$ and $2.7$.
                 The dashed line is a possible integration path that avoids all cuts. }
        \label{fig:path}
     \end{figure}

            Let us analyze the function $\sqrt{x}=f_\pm(t,\alpha,z)$ in the complex $\sqrt{x}$ plane
            depending on the values of $t\in \mathds{C}$ and $\alpha >0$.
            We denote $\sqrt{t}=a+ib$, where $a,b \geq 0$ because $\sqrt{t}$ is in the upper right quadrant.
            Because we can identify $-\sqrt{x}$ with $+\sqrt{x}$, it is sufficient to consider only one of the cuts, for example $f_-(t,\alpha,z)$.
            The endpoints of the cut correspond to $z=\pm 1$ and thus $\sqrt{x} = \pm\sqrt{t} - i\alpha$.
            The cut starts on the vertical line $\sqrt{x}=a$ with distance $\alpha$ below the point $\sqrt{t}$
            and ends on the opposite line $\sqrt{x}=-a$ with distance $\alpha$ below $-\sqrt{t}$.

            For $\alpha=0$ the cut becomes a half-circle
            \begin{equation}
               f_\pm(t,0,z) = t \left( z \pm i\sqrt{1-z^2} \right) = t\,e^{\pm i\varphi}
            \end{equation}
            with $z=\cos\varphi$.
            For $\alpha>0$ it produces more complicated shapes as shown in Fig.~\ref{fig:path}.
            When lowering $z$ from $+1$ to $-1$, the modulus of $f_-(t,\alpha,z)$ always grows;
            and as long as $\alpha < 2b$, any cut passes through the complex conjugate point $\sqrt{t^\ast} = a-ib$ when
            \begin{equation}
               z = \frac{a^2-b^2+\alpha^2/2}{a^2+b^2}\,.
            \end{equation}
            In other words, all possible cuts remain inside the colored region in Fig.~\ref{fig:path},
            which is defined by the vertical lines $\sqrt{x}=\pm a$
            and the circle with radius $|\sqrt{t}|=\sqrt{a^2+b^2}$.
            If $\alpha > b$, a contour deformation is not necessary because the cut does not cross the real axis.

     \begin{figure}[t]
     \center{
     \includegraphics[width=0.9\columnwidth]{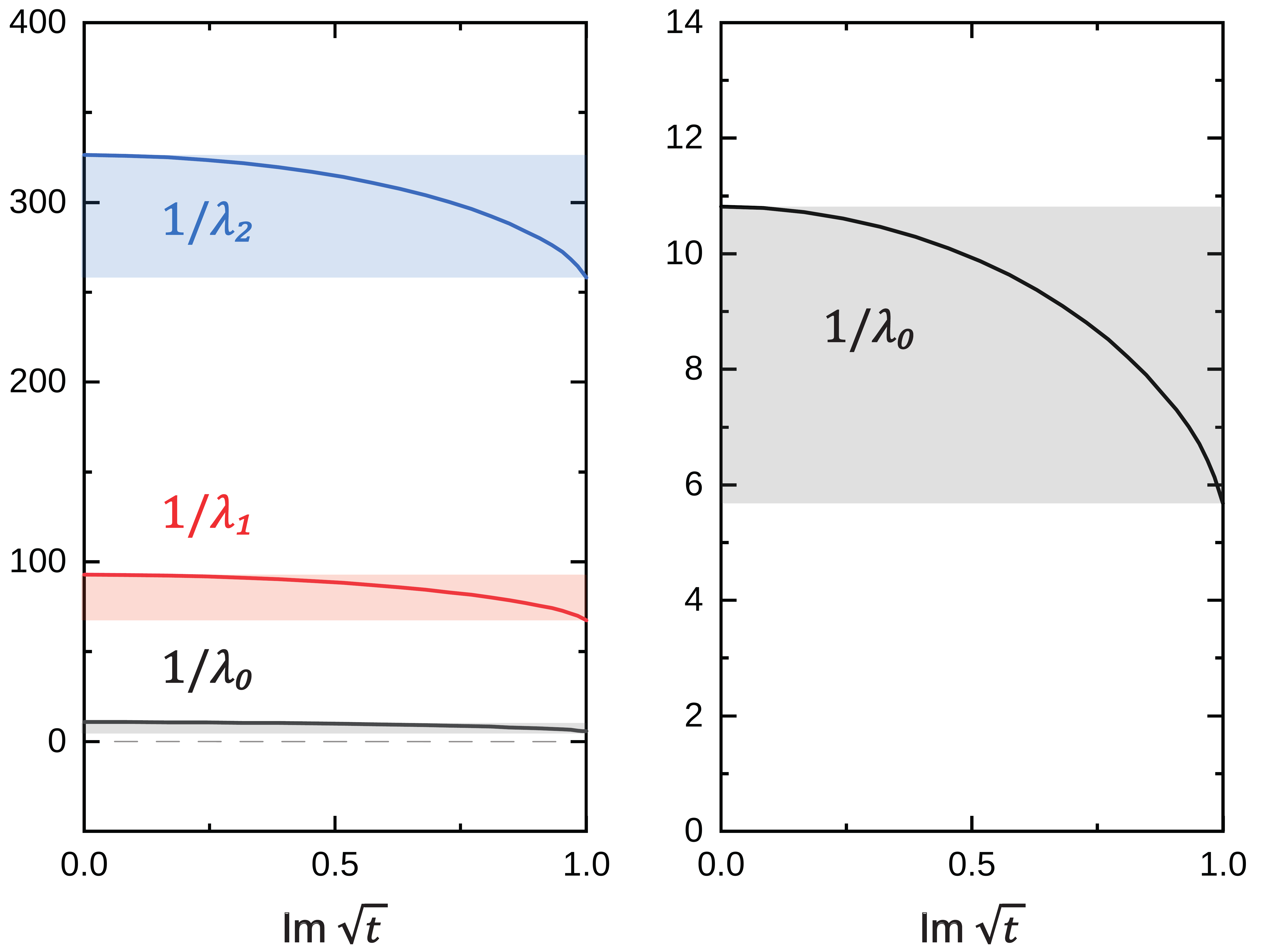}}
        \caption{Inverse eigenvalues of the Bethe-Salpeter equation for $\beta=4$ along the line $\text{Re}\sqrt{t}=0$ below threshold.}
        \label{fig:evs-real}
     \end{figure}

            Let us apply our findings to Eq.~\eqref{bse-3}.
            The propagator cut $\mC_1$ has the form discussed above with $\alpha=1$.
            If $b=\text{Im}\sqrt{t} < 1$, it does not cross the real axis and no action is required ---
            a `naive' Euclidean integration is possible.
            If $b>1$, we must deform the integration contour; at this point any possible path
            that avoids the cut is sufficient.

            The kernel cut $\mC_3$ is analogous but with $t$ replaced by $X$ and $\alpha=\beta$.
            If Eq.~\eqref{bse-3} were just an integral and not an integral equation, then for $\text{Im}\sqrt{X} < 1$
            no further steps would be necessary. However, the fact that we feed the amplitude $\Psi(X,Z,t)$ back into the
            integral on the r.h.s.
            entails that the paths for $X$ and $x$ must match, i.e., we solve
            the equation along the same path for $X$ and $x$. Therefore, at every point $\sqrt{x}$ along the path
            the kernel generates another cut $\mC_3$ that must be avoided.

            All those cuts are still confined inside regions analogous to Fig.~\ref{fig:path}, so that each point $\sqrt{x}$
            defines a corresponding cut region. To ensure that we never cross \textit{any} cut along the path,
            we must stay outside of the regions belonging to the previous points on the path.
            This means once we have reached the point $\sqrt{x}$,
            the allowed region to proceed is bounded by a vertical line and a circle.
            For $\sqrt{x}$ in the upper right quadrant we may turn left only up to a vertical line
            and we may turn right and return to the real axis not faster than on a circle.
            In other words, both $\text{Re}\sqrt{x}$ and $|\sqrt{x}|$ must never decrease along the path.

            A possible path satisfying these constraints is drawn in Fig.~\ref{fig:path}: The first section is a straight line
            connecting the origin with the point $\sqrt{t}$; the second is an arc that starts at $\sqrt{t}$ and returns to the real axis
            with increasing modulus; and the third is a straight line along the real axis up to infinity.
            In this way we can cover the entire complex plane and solve the homogeneous BSE~\eqref{bse-3}
            as well as the inhomogeneous BSE for any $\sqrt{t}\in\mathds{C}$.

            If the BS amplitude or vertex are taken fully onshell,
            we must also consider the cut $\mC_2$. Its form is analogous except that the
            point $t$ is replaced by $-(1+t)$, so that for fixed $t$ one must circumvent \textit{two} cuts in the complex $\sqrt{x}$ plane.
            This situation is discussed in Appendix~\ref{app:cd}.

     \subsection{Eigenvalues of the homogeneous equation}

           To solve the homogeneous BSE~\eqref{bse-3}, we write it as
           \begin{equation}\label{bse-4}
              \Psi(X,Z,t) = c \int \! dx \! \int \! dz\,\mK(X,Z,x,z,t)\,\Psi(x,z,t)\,.
           \end{equation}
           We pulled out the coupling $c$ since in this model it is just a constant overall factor in the kernel
           and the propagators do not depend on it because they remain at tree level.
           The contour deformation discussed above only enters in the integral measure $dx$.
           Eq.~\eqref{bse-4} can be written as a matrix-vector equation
           \begin{equation}\label{bse-5}
              \Psi_\sigma(t) = c\,\mK_{\sigma\tau}(t)\,\Psi_\tau(t)\,,
           \end{equation}
           where the matrix indices $\sigma$, $\tau$ absorb the momentum dependence in the variables $X$ and $Z$.
           In practice we calculate the eigenvalues $\lambda_i(t)$
           of the matrix $\mK(t)$, which aside from $t$ only depend on the mass ratio $\beta$
           but no longer on the coupling $c$.
           A solution of the equation and thus a pole in the scattering amplitude
           corresponds to the points $t_i$ where the condition
           \begin{equation}\label{pole-condition}
              \frac{1}{\lambda_i(t_i)} \stackrel{!}{=} c\,, \qquad i=0,1,2, \dots
           \end{equation}
           is satisfied. The respective eigenvector $\Psi_\sigma(t_i)$ is the onshell BS amplitude of that state.

     \begin{figure}[t]
     \center{
     \includegraphics[width=1\columnwidth]{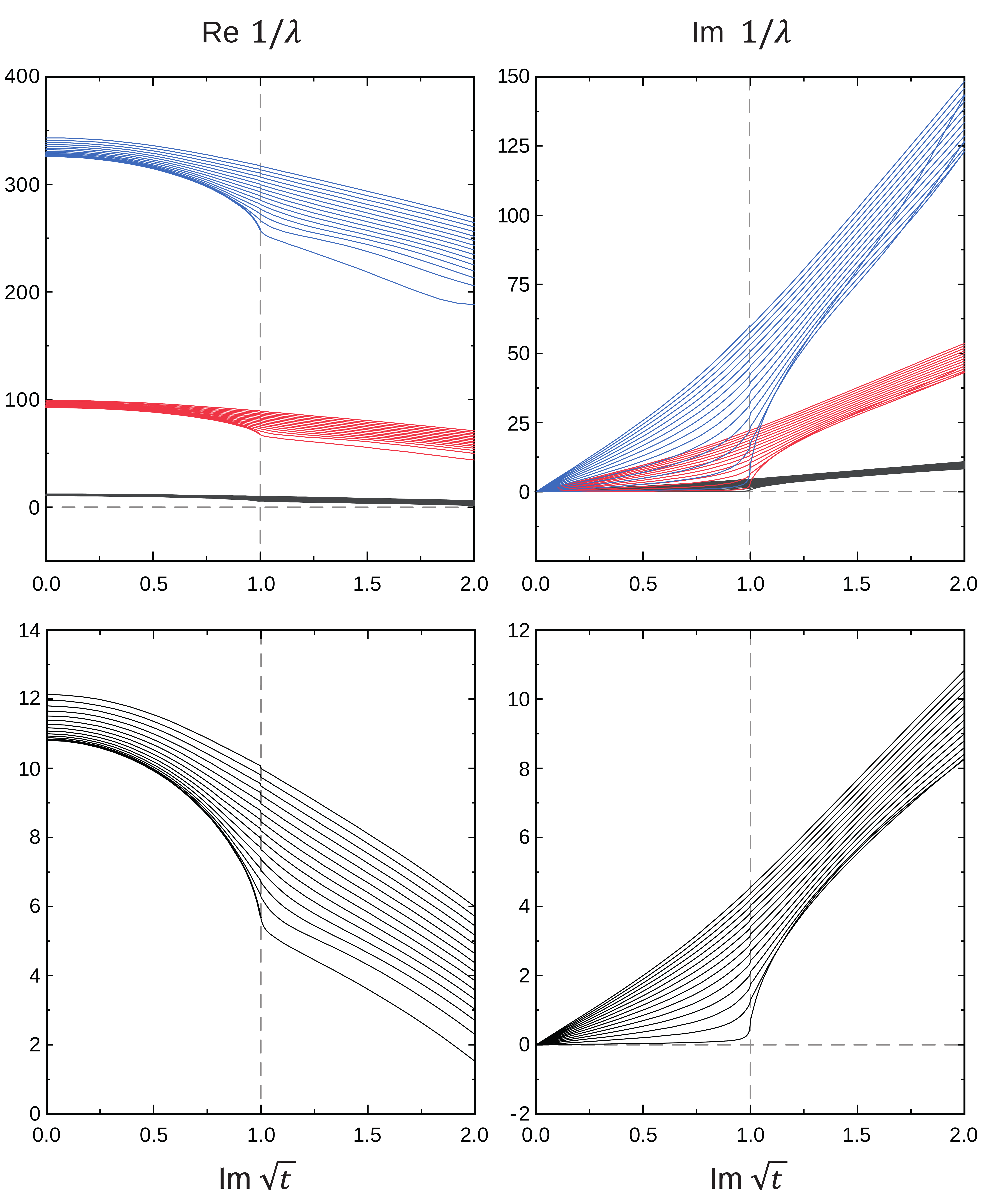}}
        \caption{Inverse eigenvalues of the Bethe-Salpeter equation for $\beta=4$
                 below and above threshold.
                 The upper panels show the first three eigenvalues and the lower panels
                 zoom in on the ground state.}
        \label{fig:evs-complex}
     \end{figure}

           The situation is illustrated in Fig.~\ref{fig:evs-real}.
           The left panel shows the first three inverse eigenvalues of $\mK(t)$
           as a function of $\text{Im}\sqrt{t}$, i.e., along the imaginary axis up to the threshold in Fig.~\ref{fig:sqrt-t-plane}.
           The right panel magnifies the eigenvalue for the ground state.
           We chose a mass ratio $\beta=4$ where the eigenvalues are well separated.
           For $c \lesssim 6$ there is no bound state.
           For $6 \lesssim c  \lesssim 11$, the ground state is bound since the condition $1/\lambda_0=c$
           can be satisfied; its mass corresponds to $\text{Im}\sqrt{t_0}=M_0/(2m)$.

           For larger couplings, the condition is still satisfied for spacelike values of $t$:
           The eigenvalues continue to drop for $P^2>0$ and the inverse eigenvalues grow for \mbox{$\sqrt{t}>0$} on the real axis.
           Therefore, with increasing coupling the pole in the scattering amplitude slides down on the imaginary axis of $\sqrt{t}$
           in Fig.~\ref{fig:sqrt-t-plane}, becomes tachyonic and continues to move along the positive real axis towards infinity.
           Recalling Sec.~\ref{sec:model}, this  signals again that for physically acceptable solutions
           the coupling of the model must be restricted to small enough values.

           If we increase the coupling further, then from Fig.~\ref{fig:evs-real} eventually also the first excited state becomes bound,
           until its mass drops to zero and becomes tachyonic; followed by the second excited state, etc.
           Hence the question: what is the nature of a state before it becomes bound?
           In particular, what happens for small couplings where
           all states, including the ground state, are
           above threshold  (here for $c \lesssim 6$)?

           One should note that the onshell BS wave function $\chi(q,P) = G_0(q,P)\,\Psi(q,P)$,
           which in quantum field theory is defined as the Fourier transform of the
           matrix element of two field operators
           between the vacuum and  the asymptotic one-particle state $|P\rangle$,
           technically only refers to the solution of Eq.~\eqref{bse-5} at a particular bound-state pole
           for real $t_i$.
           Eq.~\eqref{bse-5}, on the other hand, provides an analytic continuation of $\Psi(t)$ for general $t\in\mathds{C}$
           which follows from the scattering equation,
           so that $\Psi(t)$ is the residue of the scattering amplitude also for resonance poles in the complex plane.
           In particular, the matrix $\mK(t)$ contains the scattering information
           in the $t$ channel for given quantum numbers and
           Eq.~\eqref{pole-condition} is the general condition for a pole in the scattering matrix, irrespective of whether it
           is real or complex.

           Fig.~\ref{fig:evs-complex} shows the same three eigenvalues as before (the lower panels zoom in on the ground state),
           but now including the results above threshold obtained with the contour deformation.
           The 15 different curves correspond to evenly spaced vertical lines
           in the complex $\sqrt{t}$ plane between $0.01 < \text{Re}\sqrt{t} < 0.65$, where the smallest curve
           in each plot is the result closest to the imaginary axis.
           Very close to the axis one requires better and better numerics to get stable results above threshold;
           the curve for the ground state in Fig.~\ref{fig:evs-complex} is stable but
           artifacts in the excited states can already be seen and they grow for higher-lying states.

     \begin{figure}[t]
     \center{
     \includegraphics[width=0.85\columnwidth]{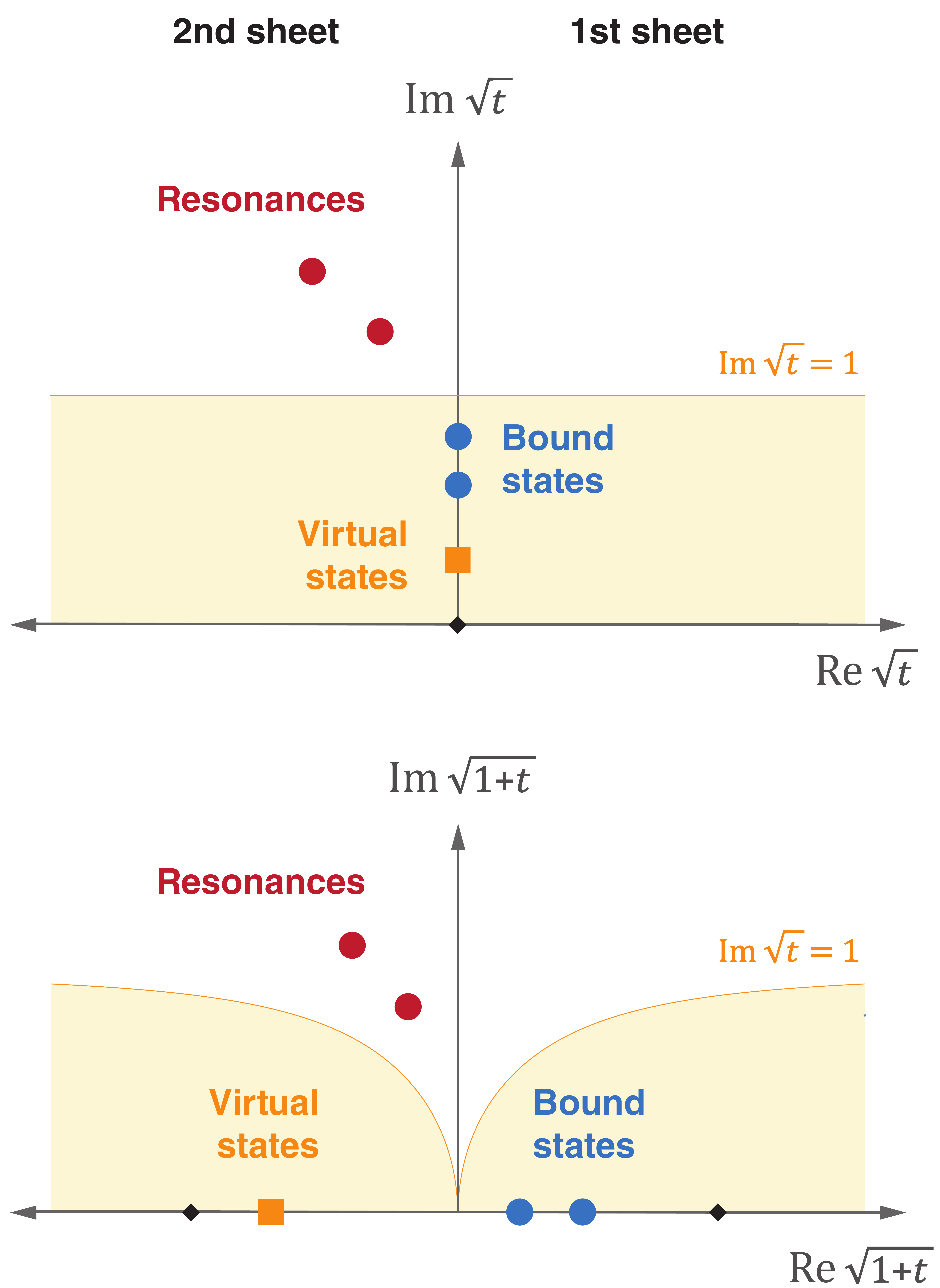}}
        \caption{\textit{Top:} Singularity structure in the complex $\sqrt{t}$ plane, with the first and second Riemann sheet plotted side by side.
                 The cut that distinguishes the two sheets now appears below threshold.
                 \textit{Bottom:} The same situation in the complex $\sqrt{1+t}$ plane where the cut no longer appears.}
        \label{fig:sqrt-1+t-plane}
     \end{figure}

           In Fig.~\ref{fig:evs-complex} one clearly sees the non-analyticity at the threshold,
           where the real and imaginary parts of all three eigenvalues have a kink
           and a branch cut opens in the imaginary parts. The condition $1/\lambda_i(t)=c$ is never satisfied
           in the complex plane and thus the first Riemann sheet is free of singularities as expected.
           In principle, however, it can be met on the second Riemann sheet, which is
           the analytic continuation of the first sheet above threshold.
           Because the coupling is real,
           the pole position is the intersection of
           \begin{equation}\label{bse-pole-positions}
               \text{Re}\,\frac{1}{\lambda_i(t)} = c \quad \text{and}\quad
               \text{Im}\,\frac{1}{\lambda_i(t)} = 0\,.
           \end{equation}

     \subsection{Continuation to the second sheet}

          To analytically continue the eigenvalues to the second Riemann sheet,
          we employ the Schlessinger point method or Resonances-via-Pad\'e (RVP) method~\cite{Schlessinger:1968},
          which has been advocated recently as a tool for locating resonance positions
          in the complex plane~\cite{Haritan:2017vvv,Tripolt:2017pzb,Tripolt:2018xeo}. It amounts to a continued fraction
          \begin{equation}
            f(z) = \frac{c_1 \qquad\qquad}{1 + \displaystyle\frac{c_2\,(z-z_1)}{1 + \displaystyle\frac{c_3\,(z-z_2)}{1 + \displaystyle\frac{c_4\,(z-z_3)}{\dots}}}}
          \end{equation}
          which is simple to implement by an iterative algorithm.
          Given $n$ input points $z_i$ with $i=1 \dots n$ and a function whose values $f(z_i)$ are known,
          one determines the $n$ coefficients $c_i$ and thereby obtains an analytic continuation
          of the original function for arbitrary values $z\in\mathds{C}$.
          The continued fraction can be recast into a standard Pad\'e form in terms of a division of two polynomials.

          The situation in the complex $\sqrt{t}$ plane is shown in the upper panel of Fig.~\ref{fig:sqrt-1+t-plane}.
          If the two Riemann sheets are plotted next to each other,
          then crossing over from the first to the second sheet is analytic above the threshold $\text{Im}\sqrt{t}=1$.
          The alignment of the two sheets (left or right) is not important because on each sheet
          the eigenvalues are analytic functions whose real parts are symmetric around the imaginary $\sqrt{t}$ axis
          and whose imaginary parts are antisymmetric.
          Bound states can appear on the first sheet along the imaginary $\sqrt{t}$ axis below the threshold and resonances
          in the complex plane of the second sheet above threshold. In addition,
          typical features of $s$-wave amplitudes are \textit{virtual} states~\cite{Gloeckle:1983,Hanhart:2014ssa,Pelaez:2015qba},
          which are poles on the imaginary $\sqrt{t}$ (or real $t$) axis on the second sheet below threshold.

          Note that with the two sheets aligned side by side there is now a cut below threshold,
          for $0 < \text{Im}\sqrt{t} < 1$, which separates the two sheets ---
          crossing over to the second sheet is smooth above threshold whereas below it is not.
          This poses a difficulty for the RVP method because a Pad\'e ansatz cannot reproduce branch cuts but only poles.
          We therefore adapt the strategy of Ref.~\cite{Hanhart:2014ssa} and
          unfold the two sheets by considering the $\sqrt{1+t}$ plane shown in the bottom panel of Fig.~\ref{fig:sqrt-1+t-plane}.
          Bound states now appear on the positive real axis,
          which eventually turns into the spacelike axis, whereas virtual states would appear as poles on the negative real axis.
          The lower half plane is again mirror-symmetric and does not provide new information.
          Importantly, there is no longer a cut in the complex $\sqrt{1+t}$ plane,
          so one can analytically continue to the second sheet also below threshold.

     \begin{figure*}[t]
     \center{
     \includegraphics[width=1.0\textwidth]{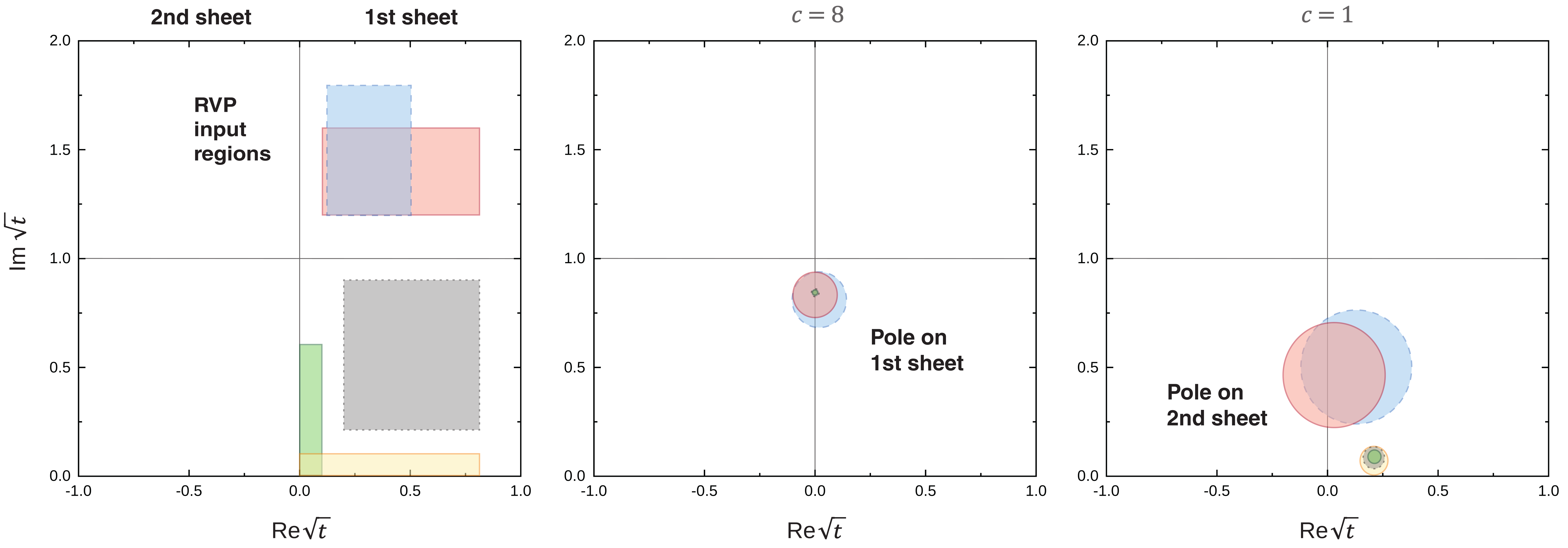}}
        \caption{Analytic continuation using the RVP method.
                 For each input region in the left panel the corresponding circles are
                 the resulting pole positions.
                 For $c=8$ one recovers the bound-state pole on the first sheet
                 whereas $c=1$ produces a pole on the second sheet below threshold.}
        \label{fig:pole-trajectories}
     \end{figure*}

          Our setup is shown in the left panel of Fig.~\ref{fig:pole-trajectories}.
          The right half plane is the first Riemann sheet in $\sqrt{t}$,
          where we calculate the BSE eigenvalues either below threshold without or above threshold with contour deformations.
          The left half plane is the second sheet.
          To determine the sensitivity of the RVP algorithm to the input region where the function is defined,
          we choose five different domains which are shown by the colored rectangles.
          Inside each rectangle we  pick $n=20,21, \dots 80$ input points randomly.
          For each rectangle and each $n$ we then determine the resulting pole positions from Eq.~\eqref{bse-pole-positions}.
          Finally we average those pole positions over $n$ to get an estimate on the sensitivity to $n$.
          For each of the five input regions, the resulting pole averages with their standard deviations
          are given by the circles in~Fig.~\ref{fig:pole-trajectories}.

          The center panel in Fig.~\ref{fig:pole-trajectories} shows the result for $\beta=4$ and $c=8$.
          This merely serves as a check because in this case we already know the result; from Fig.~\ref{fig:evs-real} and Eq.~\eqref{pole-condition}
          one can read off the bound-state pole position $\sqrt{t} \approx 0.84 \,i$.
          Indeed, the RVP method accurately reproduces this value for
          either of the three input regions below threshold, whereas the two upper regions lead to scattered poles around that value
          and thus larger circles.

          In the right panel of Fig.~\ref{fig:pole-trajectories} we show the result for a weak coupling $c=1$
          where the bound state has presumably turned into a resonance (cf.~Fig.~\ref{fig:evs-real}).
          In this case, however, we do not find resonance poles above threshold but rather below the threshold on the second sheet.
          For a given $n$ the method typically finds one pole in the $\sqrt{1+t}$ plane slightly above or below the negative real axis
          but sometimes also two.
          Transformed back to $\sqrt{t}$, the poles can dive under the cut and thus appear on the right half plane
          but they still lie on the second sheet. This is of course an artifact since the method
          is not aware of the analyticity property where
          the resulting function on the lower half plane of $\sqrt{1+t}$ must be the complex conjugate
          of that on the upper half plane; it merely performs an analytic continuation.
          Keeping this in mind, the results indeed point towards the occurrence of virtual states
          instead of resonances above the threshold.

     \begin{figure}[b]
     \center{
     \includegraphics[width=0.8\columnwidth]{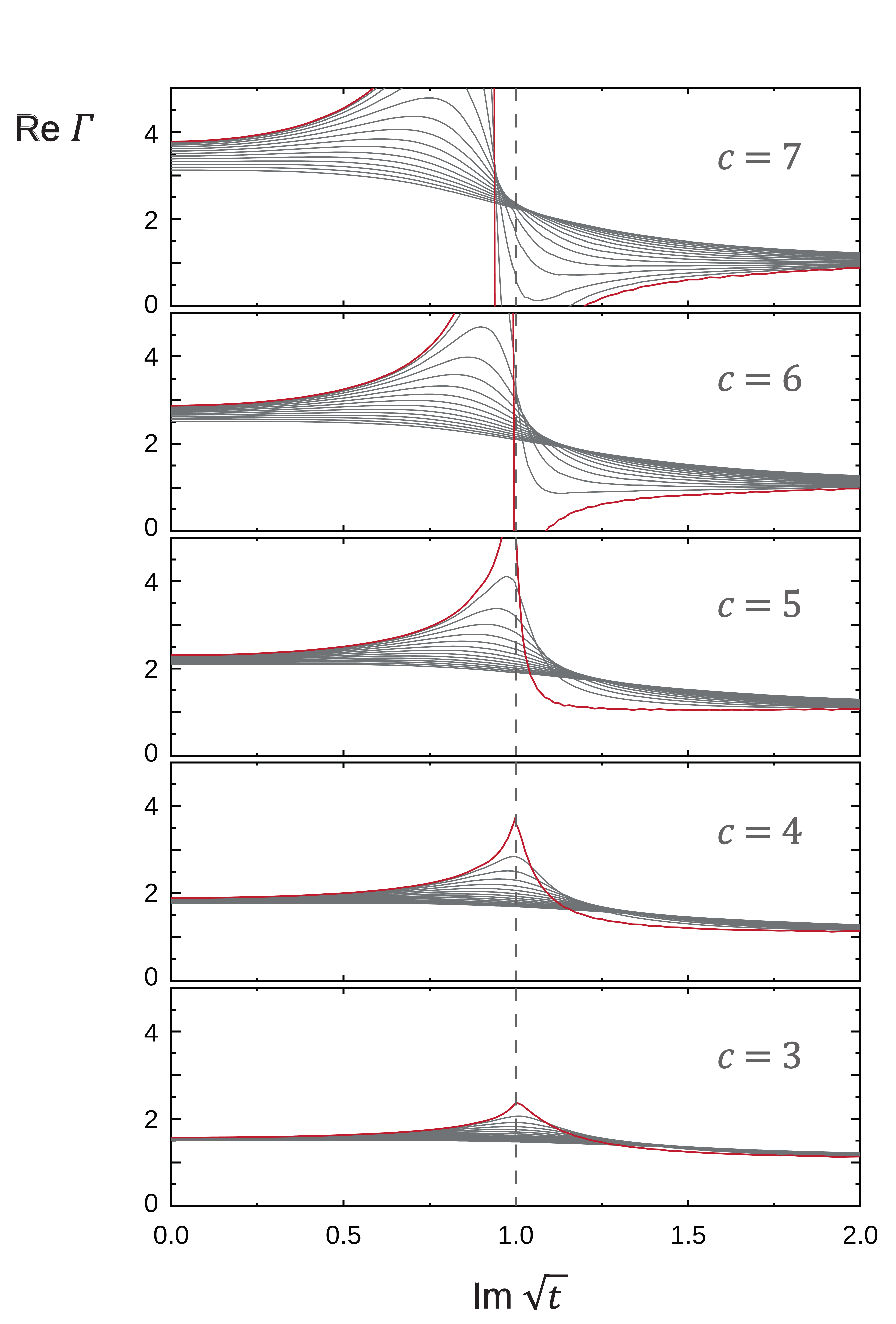}}
        \caption{Onshell vertex for different couplings. Once the bound state pole disappears,
                 only the threshold cusp survives.}
        \label{fig:no-bump}
     \end{figure}

          As an independent check we also solve the inhomogeneous BSE~\eqref{ibse-0} for the vertex $\Gamma(X,Z,t)$.
          Its practical form is analogous to Eqs.~(\ref{bse-4}--\ref{bse-5}) with the addition of an inhomogeneity
           \begin{equation}\label{ibse-3}
              \Gamma_\sigma(t) = \Gamma_{0,\sigma} + c\,\mK_{\sigma\tau}(t)\,\Gamma_\tau(t)\,,
           \end{equation}
           and instead of the eigenvalues of $\mK(t)$ the inhomogeneous BSE determines the vertex $\Gamma_\sigma(t)$ directly.
           Its singularity structure coincides with the singularities of the full scattering amplitude
           for quantum numbers $J^{PC} = 0^{++}$. Eq.~\eqref{ibse-3} can be solved by iteration, except
           in the vicinity of the poles where we employ matrix inversion to avoid convergence problems:
           \begin{equation}
              \Gamma_\sigma(t) = \left[ \mathds{1} - c\,\mK(t)\right]^{-1}_{\sigma\tau}\,\Gamma_{0,\tau}\,.
           \end{equation}
           Once $\Gamma(X,Z,t)$ is known, we solve the equation one more time to obtain
           the onshell vertex $\Gamma(-(1+t),0,t)$ which is a function of $t$ only, cf.~Eq.~\eqref{onshell-0}.

     \begin{figure*}[t!]
     \center{
     \includegraphics[scale=0.095]{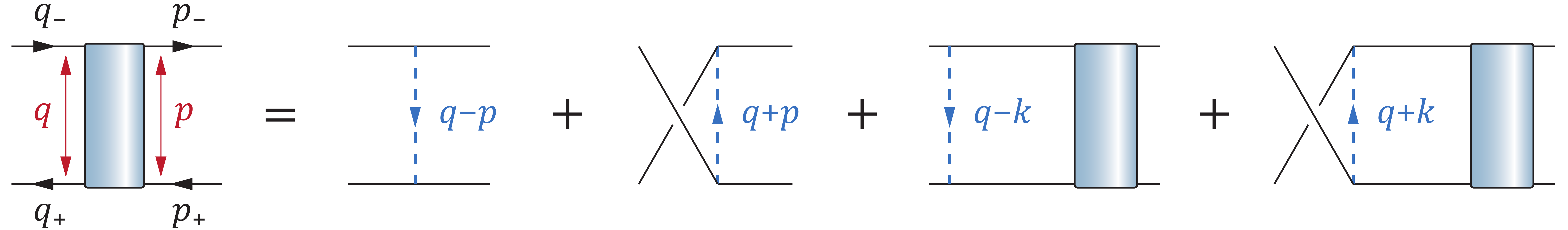}}
        \caption{Scattering equation~\eqref{sc-eq}. }
        \label{fig:bse}
     \end{figure*}

           Fig.~\ref{fig:no-bump} shows the resulting onshell vertex for $\beta=4$ and five different
           values of the coupling. The different curves correspond to the same grid as
           for the eigenvalues in Fig.~\ref{fig:evs-complex}.
           For $c=7$ and $c=6$ there is a clear bound-state pole. For $c\leq 5$ the coupling is below
           the value where bound states can occur and the ground state becomes unbound (cf.~Fig.~\ref{fig:evs-real}).
           Instead of producing a resonance bump, however,
           the pole simply disappears and what remains is just the threshold cusp.
           The cusp is the onset of the cut below threshold:
          On the $\text{Im}\sqrt{t}$ axis, the continuation of $\text{Re}\,\Gamma$ from above
          threshold becomes the function on the second sheet,
          just like for the eigenvalues in Fig.~\ref{fig:evs-complex}.
           Combined with the RVP method, the resulting pole positions from the inhomogeneous BSE are compatible
           with those in Fig.~\ref{fig:pole-trajectories} obtained from the ground-state eigenvalue of the homogeneous BSE.

           We therefore conclude that the scalar model does not produce resonances above
           threshold. Our results so far yield poles on the second sheet below threshold, which
          suggests virtual states. This poses an obstacle for analytic continuation methods since
          one has to bridge a considerable distance in the complex $\sqrt{1+t}$ plane.
          That the RVP method is well suited for finding resonance poles \textit{above} thresholds has been demonstrated in Ref.~\cite{Tripolt:2017pzb}.
          Thus, if the model \textit{did} produce clear resonance bumps, the homogeneous (or inhomogeneous) BSE
          in combination with contour deformations and the RVP method would form an adequate toolbox to determine
          the pole locations in the complex plane.

          On the other hand, there is a method that provides direct access to the second sheet:
          two-body unitarity, which follows from the scattering equation and allows
          one to calculate the scattering amplitude on the second sheet if it is known on the first sheet.
          We discuss it in more detail in Sec.~\ref{sec:2b-unitarity};
          to utilize it, we must first solve the full scattering equation in Eq.~\eqref{sc-eq-0}.

     \section{Scattering amplitude} \label{sec:sc-eq}

        \subsection{Onshell scattering amplitude} \label{sec:kinematics}

             The scattering equation for the $2 \to 2$ scattering amplitude $T(q,p,P)$
             is shown in Fig.~\ref{fig:bse} and reads explicitly:
             \begin{equation}\label{sc-eq}
                T(q,p,P) = K(q,p) + \int\limits_k K(q,k)\,G_0(k,P)\,T(k,p,P)\,.
             \end{equation}
             The ingredients are the same as before, Eqs.~(\ref{G0-0}--\ref{kernel-0}),
             except that the amplitude depends on a second relative momentum $p$
             with $p_\pm = p \pm P/2$.

             Ultimately we are only interested in the onshell scattering amplitude
             where all particles are on their mass shells, $p_\pm^2 = q_\pm^2 = -m^2$,
             which entails $p\cdot P = q\cdot P = 0$ and $p^2 = q^2 = -m^2\,(1+t)$.
             This leaves two independent variables: the total momentum transfer $t = P^2/(4m^2)$ and
             the crossing variable $\lambda = p\cdot q/m^2$.
             The latter can be written as $\lambda=(1+t)\,Y$, where the hyperspherical variable $Y=\cos\theta$
             is the cosine of the scattering angle $\theta$ in the CM frame.
             In the following we denote the onshell scattering amplitude by $T_\text{on}(t,Y)$.

\begin{figure}[t]
     \center{
     \includegraphics[width=0.85\columnwidth]{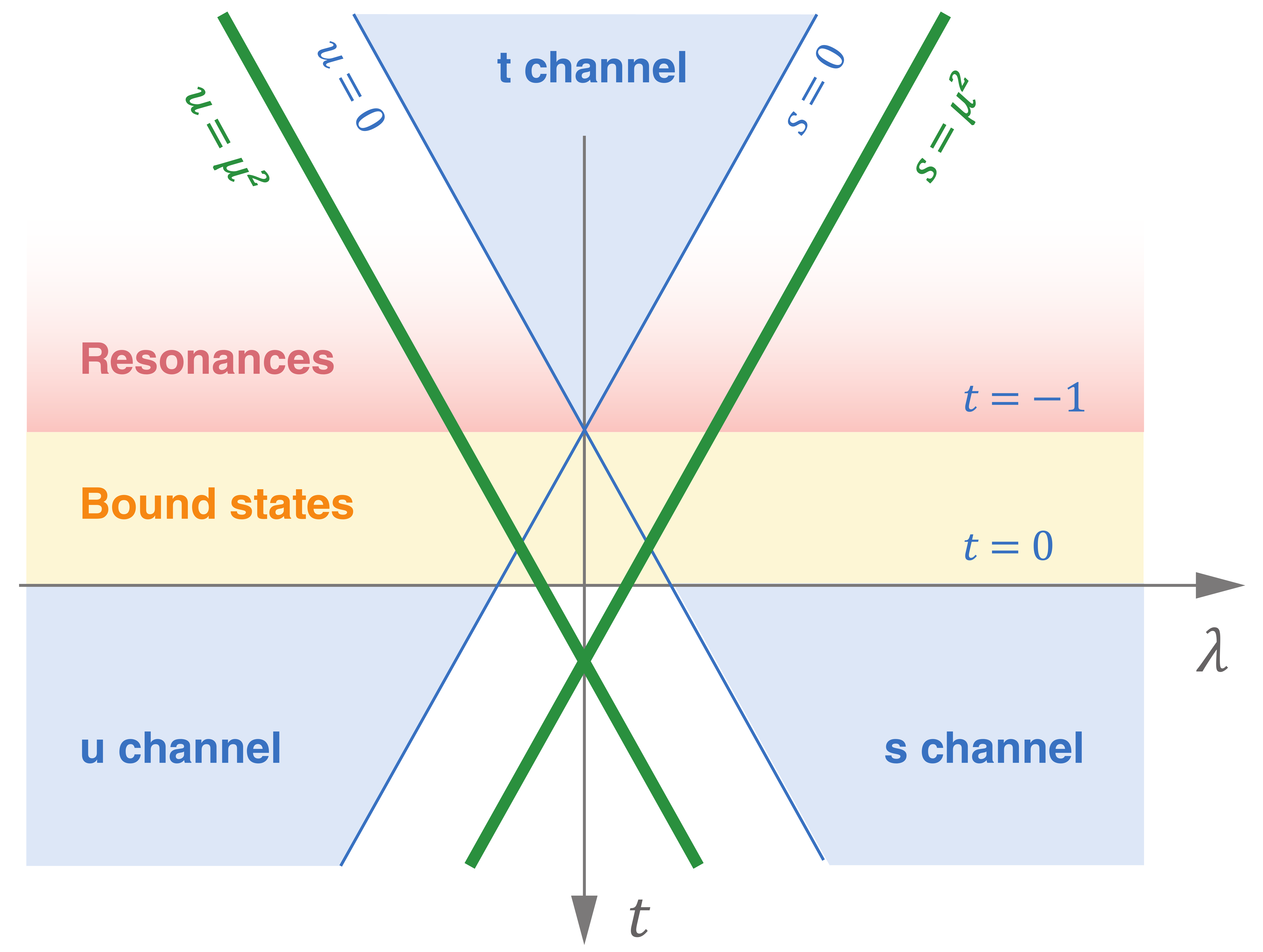}}
        \caption{Mandelstam plane for the onshell scattering amplitude in the variables $t$ and $\lambda$. }
        \label{fig:mandelstam}
     \end{figure}

             In Fig.~\ref{fig:mandelstam} we illustrate the singularity structure of $T_\text{on}(t,Y)$
             in the Mandelstam plane of the variables $t$ and $\lambda$.
             The bound states, resonance poles and $t$-channel cuts generated in the nonperturbative
             solution of the scattering equation are identical to those obtained with the (in-)homogeneous BSEs.
             They appear at fixed $t$ and are independent of $\lambda$, so they form horizontal lines in the Mandelstam plane.
             The physical $t$ channel opens above threshold $t<-1$. 

             By contrast, the singularities that depend on $\lambda$ are purely perturbative.
             On the one hand, the exchange kernel $K(q,p)$ has poles in the $s$ and $u$ channels.
             The Mandelstam variables $s$ and $u$ are given by
             \begin{equation}\label{mandelstam-variables}
                \left\{ \begin{array}{c} s \\ u \end{array}\right\} = -(q \mp p)^2 = 2m^2 \left( 1+t \pm \lambda \right)
             \end{equation}
             and thus $\lambda = (s-u)/(4m^2)$.
             The exchange poles appear at $s=\mu^2$ or $u=\mu^2$, which corresponds to
             \begin{equation}\label{poles-enter}
                \pm \lambda = \frac{\beta^2}{2} - (1+t)
             \end{equation}
             as shown in Fig.~\ref{fig:mandelstam}.
             On the other hand, by expanding the scattering equation into a ladder series,
             each perturbative loop diagram produces further cuts if $s>4\mu^2$ or $u>4\mu^2$
             and therefore
             \begin{equation}
                \pm\lambda > 2\beta^2 - (1+t)\,.
             \end{equation}
             The boundaries of the three channels
             are the lines with $s=0$ and $u=0$ corresponding to $\lambda = \mp(1+t)$
             and therefore $Y = \mp 1$.

             Because $T_\text{on}(t,Y)$ is invariant under the transformations $p\to -p$ or $q\to -q$,
             the scattering amplitude is symmetric in the crossing variable $\lambda$
             and thus can depend on $Y$ only quadratically.
             The $Y$ dependence can be absorbed in a partial-wave expansion:
             \begin{equation}\label{pw-expansion}
                T_\text{on}(t,Y) = \sum_{l=0}^\infty (2l+1)\,f_l(t)\,P_l(Y)\,,
             \end{equation}
             where $P_l(Y)$ are the Legendre polynomials and inside the $t$-channel region one has $-1 < Y < 1$.
             Because the dependence on $Y$ is quadratic, only even partial waves $l=0,2,4 \dots$ survive.

             If $\beta$ is large enough and the exchange-particle poles are far away
             from the $t$-channel region, the $\lambda$ dependence is small and the partial-wave expansion
             converges rapidly. Moreover, the exchange poles only appear in the inhomogeneous term
             of Eq.~\eqref{sc-eq}, so they drop out in the difference $T_\text{on}-K_\text{on}$ whose remaining $\lambda$-dependent singularities
             are the cuts in the $s$ and $u$ channels.
             Thus, the only \textit{nonperturbative} singularities of the scattering amplitude
             are those in the variable $t$.

        \subsection{Two-body unitarity} \label{sec:2b-unitarity}

             The scattering equation provides a direct way to access the second Riemann sheet
             via two-body unitarity, see e.g. Ref.~\cite{Gribov:2009zz} for a detailed discussion.
             If we take the inverse of the scattering equation, $T^{-1} = K^{-1} - G_0$,
             and subtract it for two different kinematical configurations we obtain
             \begin{equation}
                T_+^{-1} - T_-^{-1} = (K_+^{-1} - K_-^{-1}) - (G_{0+} - G_{0-})\,.
             \end{equation}
             Multiplying with $T_+$ from the left and $T_-$ from the right yields
             \begin{equation}\label{unitarity}
             \begin{split}
                T_+ - T_- & = T_+\,(G_{0+} - G_{0-})\,T_- \\
                          &- T_+\,(K_+^{-1} - K_-^{-1})\,T_-\,.
             \end{split}
             \end{equation}
             Using $G = G_0 + G_0\,K\,G = G_0 + G_0\,T\,G_0$,
             the second term on the r.h.s. above can be rearranged as
             \begin{equation}
             \begin{split}
                 (\dots) &= T_+\,K_+^{-1} \,(K_+-K_-)\,K_-^{-1}\, T_- \\
                         &= G_{0+}^{-1} \,G_+\,(K_+-K_-)\,G_- G_{0-}^{-1}\,.
             \end{split}
             \end{equation}

             If $T_+$ and $T_-$ are the scattering amplitudes along the $t$-channel cut ($t<-1$) with $t_\pm = t \pm i\epsilon$
             slightly displaced on either side, then the difference of the kernels drops out
             because the ladder kernel $K$ does not depend on $t$.
             The difference of the tree-level propagators 
             forces the scattering amplitudes inside the loop integral onto the mass shell and
             results in the unitarity relation
             \begin{equation}
                \text{Im}\,T_\text{on}(t,Y) = \frac{\tau(t)}{4\pi} \int\limits_{-1}^1 dy \int\limits_0^{2\pi} d\psi\,T_\text{on}(t,\Upsilon)\,T_\text{on}^\ast(t,y)
             \end{equation}
             where
             \begin{equation}
             \begin{split}
                \Upsilon &= y\,Y+\sqrt{1-y^2}\sqrt{1-Y^2}\,\cos\psi\,, \\
                \tau(t) &= \frac{1}{16\pi}\sqrt{\frac{1+t}{t}}\,.
             \end{split}
             \end{equation}

             From the knowledge of the discontinuity along the cut one therefore arrives at a relation between
             the onshell scattering amplitude on the first and second sheet:
             \begin{equation}\label{unitarity-1-2}
                T_\text{II}(t,Y) = T_\text{I}(t,Y) +   \frac{i\tau(t)}{2\pi} \int\limits_{-1}^1  dy  \int\limits_0^{2\pi}  d\psi\,T_\text{II}(t,\Upsilon)\,T_\text{I}(t,y)\,,
             \end{equation}
             which is an inhomogeneous integral equation that determines the amplitude $T_\text{II}$
             on the second sheet once $T_\text{I}$ is known.
             Inserting the partial-wave decomposition~\eqref{pw-expansion}, it becomes the algebraic relation
             \begin{equation}\label{2nd-sheet}
                 f_l(t)_\text{II} = \frac{f_l(t)_\text{I}}{1-2i\tau(t)\,f_l(t)_\text{I}}\,,
             \end{equation}
             where $f_l(t)_\text{I}$ and $f_l(t)_\text{II}$ denote the partial waves on the first and second sheet.
             Thus, the poles on the second sheet correspond to the zeros of the denominator on the first sheet:
             \begin{equation}
                f_l(t)_\text{I} \stackrel{!}{=} \frac{1}{2i\tau(t)}\,.
             \end{equation}

             Once the scattering amplitude is solved from its scattering equation~\eqref{sc-eq},
             unitarity is therefore automatic and one has direct access to the second Riemann sheet.

        \subsection{Half-offshell amplitude}\label{sec:half-offshell-amp}

             There is one complication that remains to be addressed.
             Because the internal momenta $k_\pm$ in Eq.~\eqref{sc-eq}
             are sampled in offshell kinematics, the scattering equation
             does not provide a self-consistent relation for the onshell amplitude
             but only for its half-offshell counterpart.
             Therefore, we must first solve the equation for the half-offshell amplitude;
             once completed, we perform `one more iteration' to arrive at the scattering amplitude for onshell external kinematics.

             For the half-offshell amplitude we relax the conditions $q_\pm^2 = -m^2$
             so that $q\cdot P$ and $q^2$ remain general. As a consequence, the amplitude
             now depends on four independent variables.
             On kinematical grounds the half-offshell amplitude is similar to onshell Compton scattering
             with two virtual photons, so we can use the same Lorentz-invariant momentum variables to analyze it~\cite{Eichmann:2018ytt}:\footnote{The
             kinematical analogue of Fig.~\ref{fig:bse} is the annihilation process $N\bar{N} \to \gamma^\ast \gamma^\ast$. To compare
             with the notation in Ref.~\cite{Eichmann:2018ytt},
             replace the momenta $\{ q, p, P \}$ with $\{ \Sigma, -p, -\Delta\}$.}
             \begin{equation}\label{mom-variables}
             \begin{split}
                \eta_+ &= \frac{1}{m^2}\left( q^2 + \frac{P^2}{4}\right) = \frac{q_+^2+q_-^2}{2m^2}, \\
                \eta_- &= \frac{1}{m^2}\left( q^2 - \frac{P^2}{4}\right) =  \frac{q_+ \cdot q_-}{m^2}, \\
                \omega &= -\frac{q\cdot P}{m^2} = -\frac{q_+^2-q_-^2}{2m^2}, \\[1mm]
                \lambda &= \frac{p\cdot q}{m^2} = \frac{p\cdot q_\pm}{m^2}\,.
             \end{split}
             \end{equation}
             The amplitude then depends on four variables $\eta_+$, $\eta_-$, $\omega$ and $\lambda$,
             where $\omega$ and $\lambda$ can again only enter quadratically.

             In analogy to~\eqref{frame-hyperspherical-0}, we alternatively
             use the hyperspherical variables $t$, $X$, $Y$, $Z$ defined by
             \begin{equation}\label{frame-hyperspherical}
             \begin{split}
                P &= -2m\sqrt{t} \left[ \begin{array}{c} 0 \\ 0 \\ 0 \\ 1 \end{array}\right], \quad
                p = -im \sqrt{1+t}  \left[ \begin{array}{c} 0 \\ 0 \\ 1 \\ 0 \end{array}\right], \\
                q &= m\sqrt{X} \left[ \begin{array}{l} 0 \\ \sqrt{1-Z^2}\,\sqrt{1-Y^2} \\ \sqrt{1-Z^2} \,Y \\ Z \end{array}\right], \\
                k &= m\sqrt{x} \left[ \begin{array}{l} \sqrt{1-z^2}\,\sqrt{1-y^2}\,\sin\psi \\ \sqrt{1-z^2}\,\sqrt{1-y^2}\,\cos\psi \\ \sqrt{1-z^2} \,y \\ z \end{array}\right],
             \end{split}
             \end{equation}
             where we attached minus signs to $P$ and $p$ to comply with the Compton scattering kinematics.
             This corresponds to the Lorentz-invariant definitions
             \begin{equation}\label{omega-lambda}
             \begin{split}
                \eta_\pm &= X\pm t \,, \\
                \omega &= 2\sqrt{tX}\,Z  \,, \\
                \lambda &= -i\sqrt{(1+t)\,X}\,\sqrt{1-Z^2}\,Y
             \end{split}
             \end{equation}
             and likewise for the internal variables $x$, $z$ and $\psi$ by taking dot products with the loop momentum $k^\mu$.

     \begin{figure}[t]
     \center{
     \includegraphics[width=1\columnwidth]{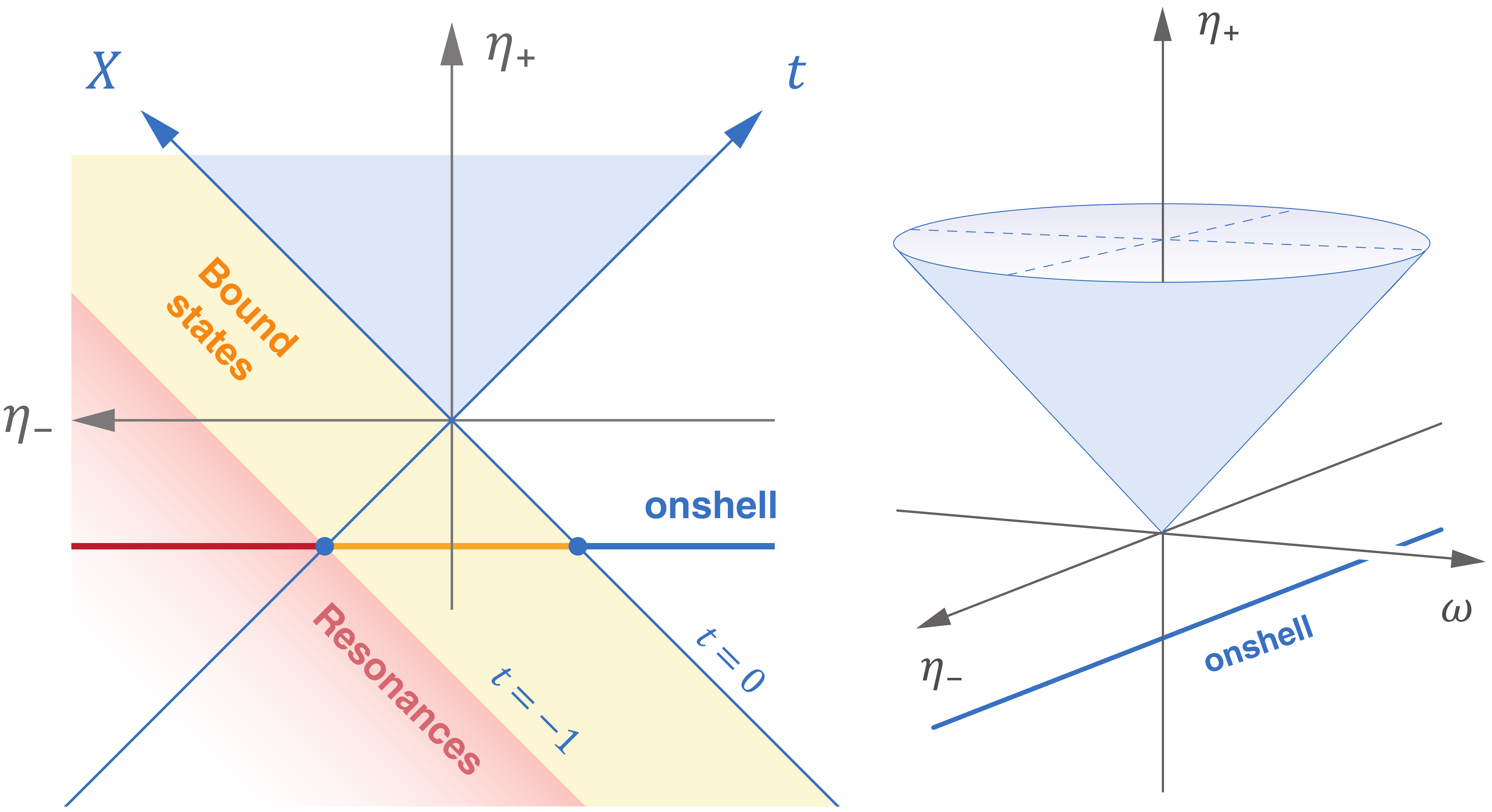}}
        \caption{Kinematic domain of the half-offshell scattering amplitude for
                 $\omega=0$ (\textit{left}) and $\omega\neq 0$ (\textit{right}).}
        \label{fig:phasespace}
     \end{figure}

             The left panel of Fig.~\ref{fig:phasespace} shows the resulting kinematic domain in $\eta_+$ and $\eta_-$ for $\omega=0$.
             As before, bound states and resonances appear at fixed $t$ below and above threshold $t=-1$, respectively.
             At a given $t$, the domain of the scattering equation where it is solved self-consistently is
             \begin{equation}\label{hyp-var}
                 X > 0\,, \quad
                 -1 < Z < 1\,, \quad
                 -1 < Y < 1\,.
             \end{equation}
             The onshell scattering amplitude $T_\text{on}(t,Y)$ corresponds to $X=-(1+t)$ and $Z=0$ and thus
             \begin{equation} \renewcommand{\arraystretch}{1.2}
             \begin{array}{rl}
                \eta_+&=-1\,, \\
                \eta_- &= -(1+2t)\,,
             \end{array}\quad
             \begin{array}{rl}
                \omega &= 0\,, \\
                \lambda &=(1+t)\,Y\,.
             \end{array}
             \end{equation}
             The onshell Mandelstam plane is then the line $\eta_+=-1$ in Fig.~\ref{fig:phasespace}.
             In the half-offshell case the Mandelstam variables $s$ and $u$ become
             \begin{equation}
                \left\{ \begin{array}{c} s \\ u \end{array}\right\} = -(q \mp p)^2 = m^2 \left( 1- \eta_- \pm 2\lambda \right)
             \end{equation}
             and the Mandelstam plane is expressed through the variables $\eta_-$ and $\lambda$.
             The exchange-particle poles correspond to $\eta_- = 1-\beta^2 \pm 2 \lambda$;
             for $\lambda=0$, the intersection of the two poles at $s=u=\mu^2$ (cf.~Fig.~\ref{fig:mandelstam})
             would form a vertical line $\eta_- = 1-\beta^2$ in Fig.~\ref{fig:phasespace}.

             If we also switch on the remaining variable $\omega$, then for $t>0$ the integration domain forms a
             cone around the $\eta_+$ axis, which is shown in the right panel of Fig.~\ref{fig:phasespace}.
             This is so because
             from Eq.~\eqref{hyp-var} one has
             \begin{equation}\label{cone-derivation}
                -\eta_+ < \eta_- < \eta_+ \quad \text{and} \quad
                 Z^2 = \frac{\omega^2}{\eta_+^2-\eta_-^2} < 1\,.
             \end{equation}
             For $t<0$ the integration domain forms a similar cone around the
             $\eta_-$ axis and with $\omega$ imaginary.
             Hence, to obtain a self-consistent solution
             one must first solve the equation inside these cones,
             and afterwards solve the equation once more by setting $X=-(1+t)$ and $Z=0$,
             where the half-offshell amplitude is integrated over
             to obtain the onshell scattering amplitude $T_\text{on}(t,\lambda)$.

             To cast the scattering equation~\eqref{sc-eq} into a form analogous to Eq.~\eqref{bse-3}, we
             express it in hyperspherical variables:
            \begin{equation}\label{sc-eq-3}
            \begin{split}
               &T(X,Y,Z,t) = K\left(X,-(1+t),\sqrt{1-Z^2}\,Y\right) \\
                & \;\; + \frac{m^4}{(2\pi)^4}\,\frac{1}{2} \int\limits_0^\infty dx\,x\int \limits_{-1}^1 dz\sqrt{1-z^2}\,G_0(x,z,t) \\
                            &  \;\;\quad \times \int\limits_{-1}^1 dy \int\limits_0^{2\pi} d\psi\,K(X,x,\widetilde\Omega)\,T(x,y,z,t)\,,
            \end{split}
            \end{equation}
            where $K$ and $G_0$ are the same as before,
            \begin{equation}\label{KG-3}
            \begin{split}
               K(X,x,\widetilde\Omega) &= \frac{g^2}{m^2}\,\frac{X+x+\beta^2}{(X+x+\beta^2)^2 - 4Xx\,\widetilde\Omega^2}\,, \\
               G_0(x,z,t) &= \frac{1}{m^4}\,\frac{1}{(x+t+1)^2-4xt\,z^2}\,,
            \end{split}
            \end{equation}
            except that the argument $\widetilde\Omega$ also depends on the angle $\psi$:
            \begin{equation}\label{omega}
            \begin{split}
            \widetilde\Omega &=  zZ + \sqrt{1-z^2}\sqrt{1-Z^2} \,\Upsilon\,,   \\
            \Upsilon & =  y\,Y+\sqrt{1-y^2}\sqrt{1-Y^2}\,\cos\psi \,.
            \end{split}
            \end{equation}
            The innermost $\psi$ integration is no longer trivial
            but it can still be performed analytically:
           \begin{equation}
               \frac{1}{2\pi}\int\limits_{0}^{2\pi} d\psi\,\frac{1}{1-(a+b\,\cos\psi)^2}
               = \frac{\gamma_+ + \gamma_-}{2b}\,,
           \end{equation}
           where
           \begin{equation}
             \gamma_\pm =  \frac{\delta_\pm}{\sqrt{1-\delta_\pm^2}}\,, \qquad
             \delta_\pm = \frac{b}{1\pm a}\,.
           \end{equation}
            The external kernel in the first line of Eq.~\eqref{sc-eq-3}
            follows from replacing the appropriate momentum arguments.

            The analysis of branch cuts is analogous to Sec.~\ref{sec:cd}.
            We keep the integration domains of $z$ and $y$ inside their intervals in Eq.~\eqref{hyp-var}
            and deform the integration contour in $\sqrt{x}$ only. In this way
            we cover the area $|\lambda| < |1+t| \Leftrightarrow |Y| < 1$, i.e., the region between the lines $s=0$ and $u=0$ in the Mandelstam plane.
            The cuts in the integrand are as follows:
            \begin{itemize}
            \item[(1)] The propagator poles encoded in $G_0(k,P)$ generate a cut $\sqrt{x}=\mathcal{C}_1 = f_\pm(t,1,z)$;
                       the function $f_\pm$ has been defined in Eq.~\eqref{f-cut-0}.
            \item[(2)] The exchange particle poles in $K(q,k)$ produce a cut $\sqrt{x}=\mathcal{C}_3 =f_\pm(X,\beta,\widetilde\Omega)$.
            \item[(3)] The exchange-particle poles in the $s$ and $u$ channels,
                       which are generated in the internal scattering amplitude $T(x,y,z,t)$ itself,
                       produce cuts in the same place as those in $K(k,p)$, cf.~Eq.~\eqref{kernel-0},
                       namely at $\sqrt{x}=\mathcal{C}_2 =f_\pm(-(1+t),\beta,\sqrt{1-z^2}\,y)$.
            \item[(4)] Finally, if we solve the equation once more in the final step to obtain the onshell scattering amplitude,
                       the exchange-particle poles in $K(q,k)$ produce a cut $\sqrt{x}=\mathcal{C}_4 =f_\pm(-(1+t),\beta,\sqrt{1-z^2}\,\Upsilon)$.
            \end{itemize}
            Since the third argument of $f_\pm$ always takes values in the interval $(-1,1)$, the cuts
            $\mC_2$ and $\mC_4$ lead to the same condition and do not need to be discussed separately.
            As before, the bound-state poles and cuts which are dynamically generated in the equation
            depend on $t$ only and we avoid them by discarding the imaginary axis $\text{Re}\sqrt{t}=0$.

            The essential complication here is the appearance of \textit{two} cuts, $\mC_1$ and $\mC_2$,
            which both depend on the external point $t$ and must be avoided. As a consequence, the kinematically safe region
            where no contour deformation is necessary shrinks to the intersection of the two conditions
            \begin{equation}\label{cond-0}
               \text{Im}\,\sqrt{t} < 1 \quad \text{and} \quad |\text{Im}\,\sqrt{-(1+t)}| < \beta\,.
            \end{equation}
            Moreover, it is no longer possible to cover the entire complex $\sqrt{t}$ plane
            using a contour deformation in $\sqrt{x}$ only; to do so, one would have to deform contours
            in the remaining variables $z$ and/or $y$ as well.
            The cut $\mC_3$, on the other hand, leads to the same condition as before,
            namely that the deformed path in $\sqrt{x}$ avoiding the cuts $\mC_1$ and $\mC_2$ must be chosen such that
            $\text{Re}\sqrt{x}$ and $|\sqrt{x}|$ never decrease along it.
            The details of the contour deformation procedure are given in Appendix~\ref{app:cd}.

            In practice the half-offshell scattering equation~\eqref{sc-eq-3} turns again into a matrix-vector equation
            analogous to Eqs.~\eqref{bse-5} and~\eqref{ibse-3}:
            \begin{equation}
               \mT_\sigma(t) = c\,\mK'_\sigma(t) + c\,\mK'_{\sigma\tau}(t)\,\mT_\tau(t) \,,
            \end{equation}
            where the multi-indices $\sigma$, $\tau$ encode the dependence on the variables $X$, $Y$ and $Z$.
            To avoid convergence problems in the iterative solution in the vicinity of the poles,
            we solve the equation through matrix inversion:
            \begin{equation}
               \mT_\sigma(t) = \left[ \mathds{1} - c\,\mK'(t)\right]^{-1}_{\sigma\tau}\,c\,\mK'_\tau(t)\,.
            \end{equation}

        \subsection{Results for the scattering amplitude} \label{sec:results-1}

     \begin{figure}[h!]
     \center{
     \includegraphics[width=1\columnwidth]{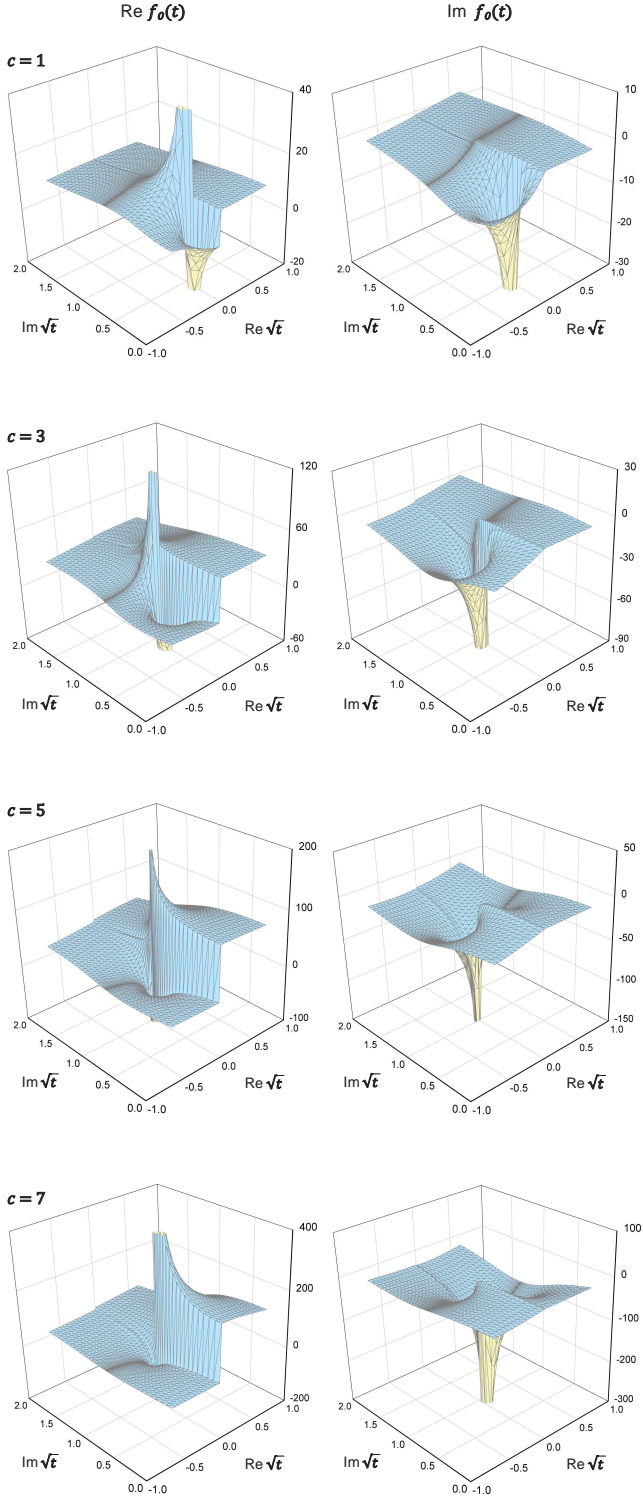}}
        \caption{Leading partial wave $f_0(t)$ in the complex $\sqrt{t}$ plane for $\beta=4$.
                 In each plot the first sheet is shown on the right ($\text{Re}\sqrt{t}>0$)
                 and the second on the left ($\text{Im}\sqrt{t}<0$), as in Fig.~\ref{fig:pole-trajectories}.
                 With increasing coupling $c$, the virtual state moves up along the imaginary axis
                 of the second sheet until it reaches the threshold and comes back on the first sheet.}
        \label{fig:2nd-sheet-3d}
     \end{figure}

     \begin{figure}[t]
     \center{
     \includegraphics[width=0.9\columnwidth]{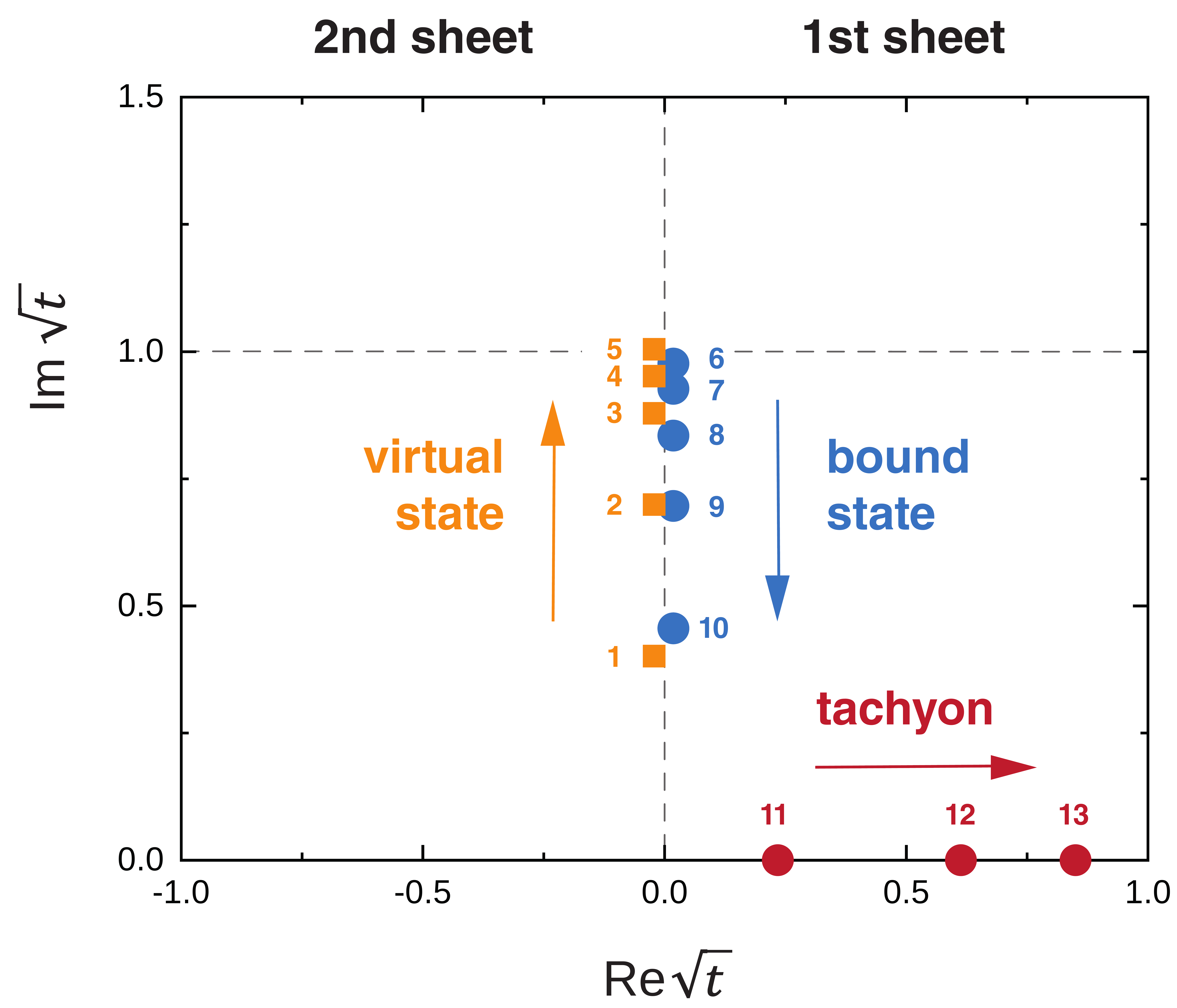}}
        \caption{Ground-state pole trajectory for $\beta=4$ as a function of the coupling
                  for $c=1,\,2,\,3\dots 13$.}
        \label{fig:pole-trajectories-2}
     \end{figure}

            Following the steps above, we first solve the half-offshell amplitude $T(X,Y,Z,t)$ from its scattering equation~\eqref{sc-eq-3}
            and obtain the onshell amplitude $T_\text{on}(t,Y)$ by solving the equation once more for $X=-(1+t)$ and $Z=0$.
            Finally we extract the partial-wave amplitudes $f_l(t)$ through
            \begin{equation}\label{pw-coeff}
               f_l(t) = \frac{1}{2} \int\limits_{-1}^1 dY\,P_l(Y)\,T_\text{on}(t,Y)
            \end{equation}
            and determine the amplitude on the second Riemann sheet from Eq.~\eqref{2nd-sheet}.

            The integration domain $Y \in [-1,1]$ is the area enclosed by the lines $s=0$ and $u=0$ in the Mandelstam plane
            of Fig.~\ref{fig:mandelstam}. In the $t$-channel region ($t<-1$) the exchange-particle poles do not enter in the integration domain,
            whereas for $t>-1$ they do appear above a certain value of $t$,
            \begin{equation}\label{tp}
               t > t_P =  \beta^2/4 - 1\,,
            \end{equation}
            which follows from Eq.~\eqref{mandelstam-variables} as the intersection of $s=\mu^2$ and $u=0$.
            As discussed above,
            the exchange-particle poles only appear in the inhomogeneous term of the scattering equation.
            If we split the partial waves into
            \begin{equation}
               f_l(t) = f_l^{(K)}(t) + f_l^{(T-K)}(t)\,,
            \end{equation}
            then the exchange poles only contribute to the first term, whose integrand
            from Eqs.~(\ref{sc-eq-3}--\ref{KG-3}) takes the form
            \begin{equation}
            \begin{split}
               K_\text{on}(t,Y) &= K(-(1+t),-(1+t),Y)  \\
                                      &= \frac{g^2}{m^2}\,\frac{1}{2(1+t)}\,\frac{B}{B^2-Y^2}
            \end{split}
            \end{equation}
            with $B = \beta^2/\left[2(1+t)\right]-1$.
            The $f_l^{(K)}(t)$ can thus be obtained analytically:
            \begin{equation}\label{exchange-poles-in-PW}
            \begin{split}
              f_0^{(K)}(t) &=  \frac{g^2}{m^2}\,\frac{1}{4(1+t)}\,\ln\frac{B+1}{B-1}\,, \\
              f_2^{(K)}(t) &=  \frac{g^2}{m^2}\,\frac{1}{4(1+t)}\left[ \frac{3B^2-1}{2}\,\ln\frac{B+1}{B-1}-3B\right] \\
            \end{split}
            \end{equation}
            and so on for higher partial waves. These expressions have poles for $t = t_P$ and branch cuts for $t > t_P$.
             The dynamically generated bound-state and resonance poles, on the other hand, only appear in the second term $f_l^{(T-K)}(t)$
             whose integrand $T_\text{on} - K_\text{on}$ is almost independent of $Y$ because
             the only singularities in that variable are the $s$- and $u$-channel cuts beginning at $t > \beta^2 - 1$.

     \begin{figure}[t]
     \center{
     \includegraphics[width=0.7\columnwidth]{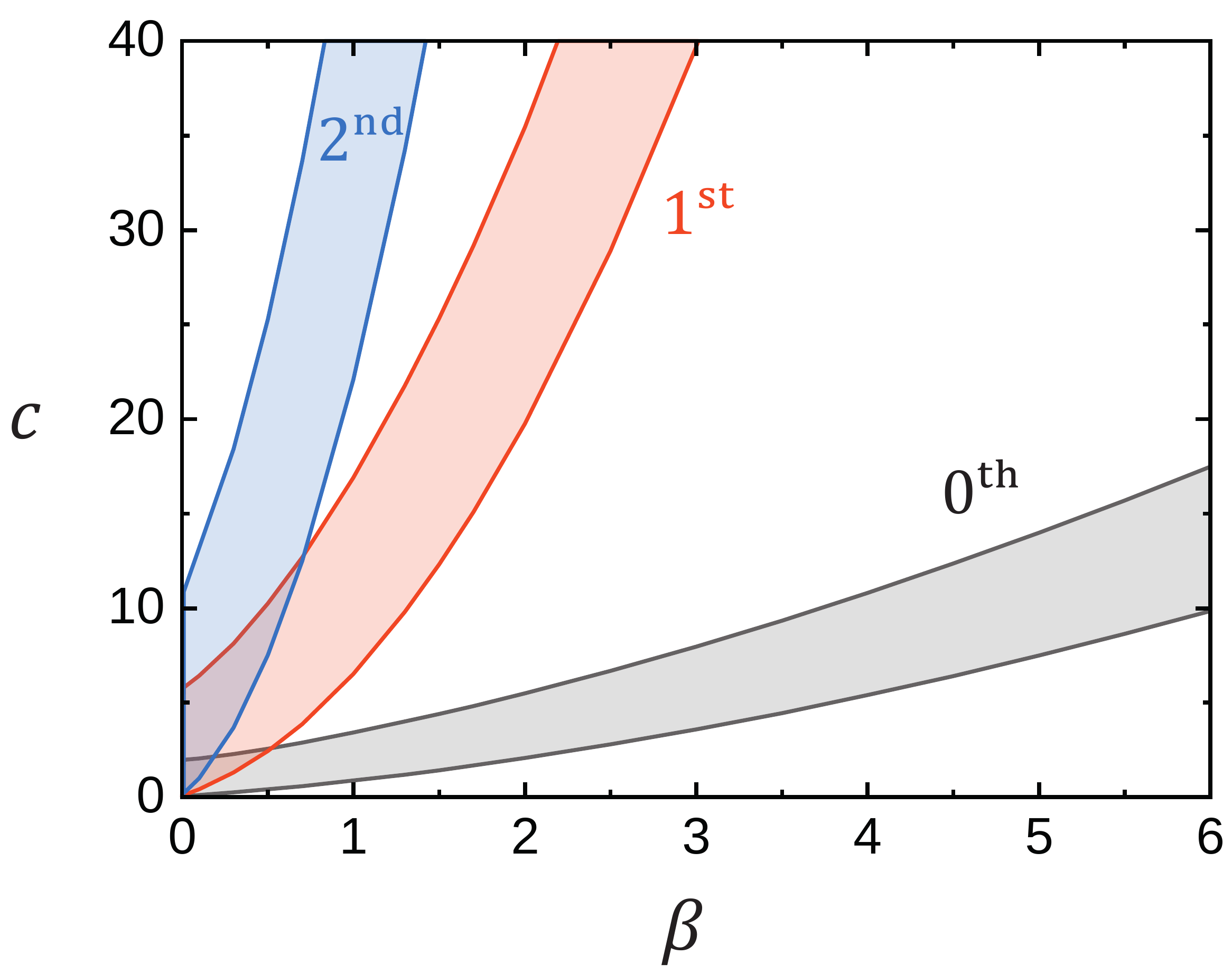}}
        \caption{Parameter space of the scalar model. The bands show the regions in $(\beta,c)$
                 where the ground state, first excited or second excited state are bound states.}
        \label{fig:parameter-space}
     \end{figure}

            Fig.~\ref{fig:2nd-sheet-3d} shows the leading partial wave $f_0(t)$ in the complex $\sqrt{t}$ plane for $\beta=4$
            and four values of the coupling $c=\{1, 3, 5, 7\}$.
            In this case the exchange poles are relatively far away from the displayed region ($t_P=3$)
            so that only the dynamical poles are visible.
            As before we plot the first Riemann sheet on the right half plane
            and the second sheet on the left.
            Recalling Fig.~\ref{fig:evs-real}, bound states
            can only appear in the interval $6 \lesssim c  \lesssim 11$. One can clearly see that below $c \sim 6$ the ground state of the system is a virtual state
            with its pole on the imaginary $\sqrt{t}$ axis (or real $t$ axis) on the second sheet.
            With increasing coupling strength $c$ it moves up towards the threshold.
            At $c\sim 6$ the pole crosses over to the first sheet, where it
            becomes a bound state, and slides down along the axis towards the origin.
            For $c \gtrsim 11$, it becomes tachyonic and
            continues on the real $\sqrt{t}$ axis.

            The resulting pole trajectory is shown in Fig.~\ref{fig:2nd-sheet-3d}.
            For $c=1$ the location of the virtual pole  is $\sqrt{t} \approx 0.40\, i$, which
            is compatible with the RVP estimate in Fig.~\ref{fig:pole-trajectories} for the two input regions above threshold.
            Thus, the homogeneous BSE in combination with the RVP method did indeed
            predict the poles of scattering amplitude  on the second sheet in the right ballpark.
            To determine the precise locations, however, we had to solve the scattering equation
            to have access to the unitarity relation~\eqref{unitarity-1-2}.

            The pattern in Fig.~\ref{fig:pole-trajectories-2} repeats itself for the excited states.
            If the coupling increases further, eventually the first excited state appears as a virtual state
            and follows the same trajectory, until it becomes bound (from Fig.~\ref{fig:evs-real} at $c \sim 66$) and then again tachyonic ($c \sim 93$),
            followed by the second excited state and so on.

            We should note that tachyonic bound states above a critical value of the coupling
            are not necessarily tied to the vacuum instability of the cubic interaction but they can also appear as truncation artifacts.
            The conditions for tachyonic poles are that (a) the eigenvalues of the system are
            monotonically decreasing functions of $t$ when passing from the timelike ($t<0$) to the spacelike side ($t>0$),
            and (b) that the coupling, when pulled out of the kernel, can be tuned independently without affecting the eigenvalues $\lambda_i(t)$.
            The first condition is a generic feature of BSEs since the eigenvalues of $\mK(t)$
            always inherit the falloff with $t$ encoded in the propagators;
            it means that in Fig.~\ref{fig:evs-real} the inverse eigenvalues continue to grow when extending the plots to the spacelike region on the left.
            Then, for a large enough coupling the condition~\eqref{pole-condition} can always be met at some value $t>0$.
            The second feature, on the other hand, is special to our model
            in the sense that there is only one overall coupling that can be pulled out of the kernel.
            In general, when solving the DSEs for the propagators they also become functions of $c$
            and so does the kernel beyond simple truncations.
            Increasing the coupling typically increases the self-energies and thus the inverse BSE eigenvalues,
            so that Eq.~\eqref{pole-condition} may no longer have a solution for spacelike $t$.
            This is effectively what happens in QCD, where the coupling $\alpha_s$ in the BSE kernel cannot be dialled independently of the quark propagator;
            instead, the strong dependence of the propagator on $\alpha_s$ is manifest in dynamical chiral symmetry breaking
            and also changes its singularity structure in the process.

            The results discussed so far for $\beta = 4$ are generic and qualitatively also hold for different values of $\beta$.
            This can be inferred from Fig.~\ref{fig:parameter-space}, where the bound-state regions in the $(\beta,c)$ plane are plotted for the lowest three
            states of the system. Inside the lowest band
            the ground state is bound; for smaller values of $c$ it is a virtual state and for larger values it becomes tachyonic.
            The same pattern repeats itself
            for the excited states.

\begin{figure}[t]
     \center{
     \includegraphics[width=1\columnwidth]{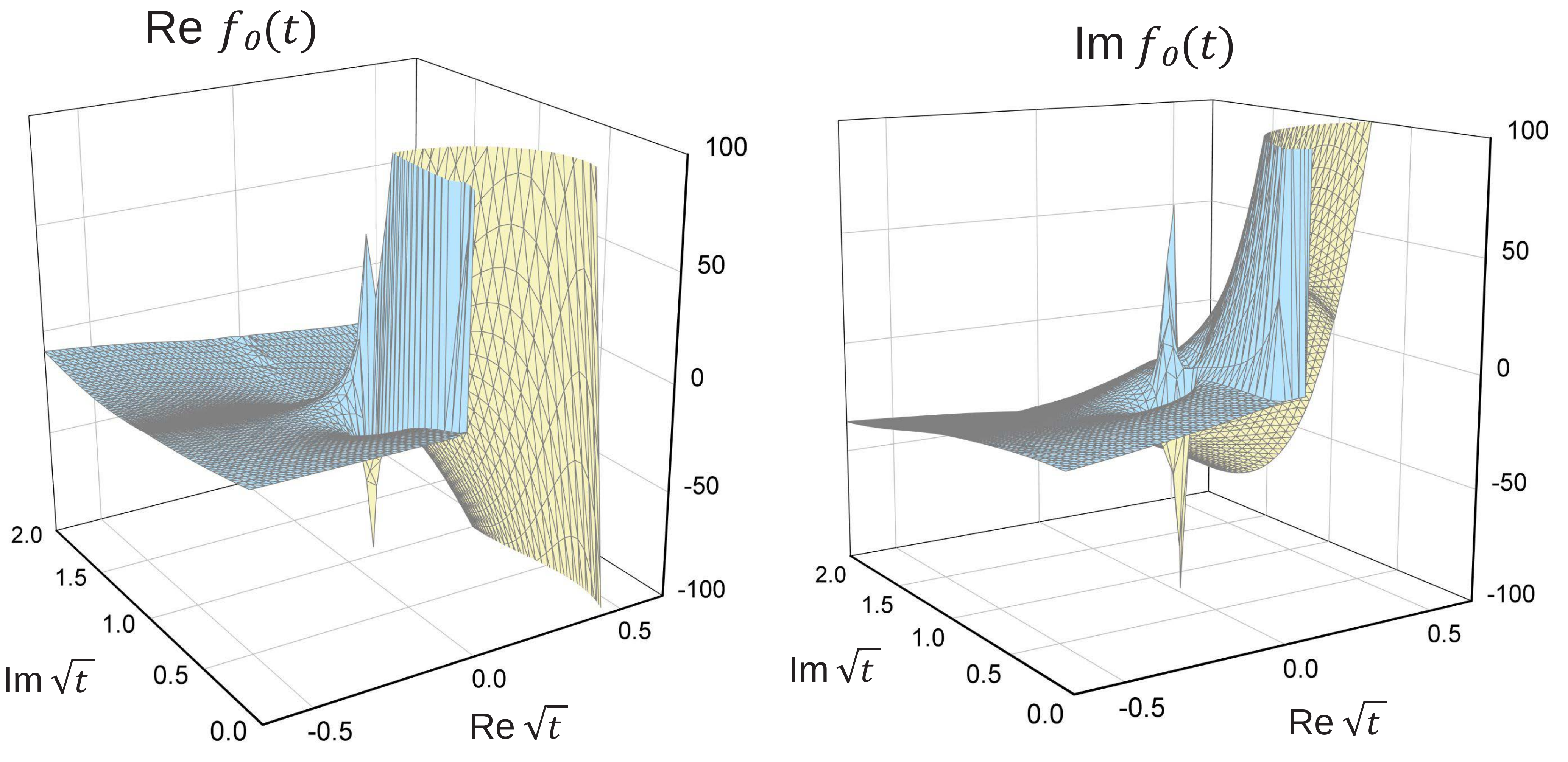}}
        \caption{Same as in Fig.~\ref{fig:2nd-sheet-3d} but for $\beta=2$ and $c=12$. The broad structure is the exchange-particle pole at $t=0$
                 and the narrow peak is the first excited state on the second sheet. }
        \label{fig:2nd-sheet-3d-2}
     \end{figure}

            For smaller values of $\beta$
            eventually also the exchange-particle poles
            from Eq.~\eqref{exchange-poles-in-PW} become important. Fig.~\ref{fig:2nd-sheet-3d-2}
            exemplifies $f_0(t)$ for $\beta=2$ and $c=12$, which lies between
            the first and second band in Fig.~\ref{fig:parameter-space}. In this case
            the ground state has become tachyonic and is no longer visible in the plot;
            instead, the large structure on the first sheet is the exchange-particle pole at $t_P=\beta^2/4-1=0$ from Eq.~\eqref{tp}.
            One can also see the first excited state on the second sheet, which has not yet become bound and is still a virtual state below threshold.

            In the limit $\beta=0$ of a massless exchange particle, which is the Wick-Cutkosky model,
            all `normal' eigenvalues in Fig.~\ref{fig:evs-real} would end at the threshold so that
            $1/\lambda_i(t=-1)=0$.
            In this case there are only bound states (and tachyons for large enough couplings)
            but no virtual states.
            The states that do not satisfy this property, i.e. $1/\lambda_i(t = -1) \neq 0$, are the so-called anomalous states~\cite{Wick:1954eu,Cutkosky:1954ru,Nakanishi:1969ph,Ahlig:1998qf}.
            To investigate what becomes of them when they cross the threshold remains the subject of future work.

     \section{Conclusions and outlook} \label{sec:summary}

          We have investigated contour deformations as a tool to extract resonance properties from
          bound-state and scattering equations. As an example we have solved the homogeneous and inhomogeneous Bethe-Salpeter equation
          for a scalar model with a ladder-exchange interaction and we calculated the $2 \to 2$ scattering amplitude from its scattering equation.
          We find that the model does not produce  resonances above threshold but rather virtual states on the second Riemann sheet.

          Our analysis was carried out in several steps.
          First, we employed contour deformations to access the kinematic regions
          associated with amplitudes in `Minkowski space'. We pointed out that there is no intrinsic difference between Minkowski and
          Euclidean space approaches; in both cases one needs to deform integration contours and the result must be the same.
          We used a Euclidean metric, which allowed us to formulate the four-dimensional integrals in a manifestly
          Lorentz-invariant way, and performed the contour deformations in the radial variable.

          We demonstrated that the resonance information is in principle already contained in the homogeneous Bethe-Salpeter equation,
          because it determines an analytic continuation of the Bethe-Salpeter amplitude which in quantum field theory is
          only defined at a particular (real) bound-state pole. To extract that information, however, one needs analytic continuation methods.
          We employed the Schlessinger point or Resonances-via-Pad\'e method, which is well-suited for situations with
          resonances above threshold. However, the scalar model we investigated here produces virtual states below threshold
          whose pole locations are more difficult to determine accurately by analytic continuation than nearby resonances.

          To address this, we solved the half-offshell scattering equations for the $2 \to 2$ scattering amplitude,
          from where we extracted the onshell scattering amplitude and employed the two-body unitarity relation
          which provides direct access to the second Riemann sheet. This allows us
          to determine the pole positions precisely and confirms that the resonance poles of the model
          are indeed virtual states below the threshold.

          Our analysis leaves room for several future investigations.
          One can study scattering amplitudes with
          complex propagator poles, the nature of anomalous states,
          or extend the approach to three-body systems.
          Our determination of the cut structure does not depend on the scalar nature of the system
          and can be applied without changes to the scattering of particles with any spin,
          such as $NN$ or $N\pi$ scattering, as long as self-energies can be neglected.
          Moreover, contour deformations provide a general way to overcome
          singularity restrictions for example in QCD, which can help to access properties of resonances
          but also highly excited states, form factors in the far spacelike and timelike regions,
          or general matrix elements in kinematic regions where a naive Euclidean integration contour is no longer applicable.

         \subsection*{Acknowledgements}

         This work was funded by the Funda\c c\~ao para a Ci\^encia e a Tecnologia (FCT) under Grants IF/00898/2015 and UID/FIS/00777/2013.

     \begin{figure}[t]
     \center{
     \includegraphics[width=0.9\columnwidth]{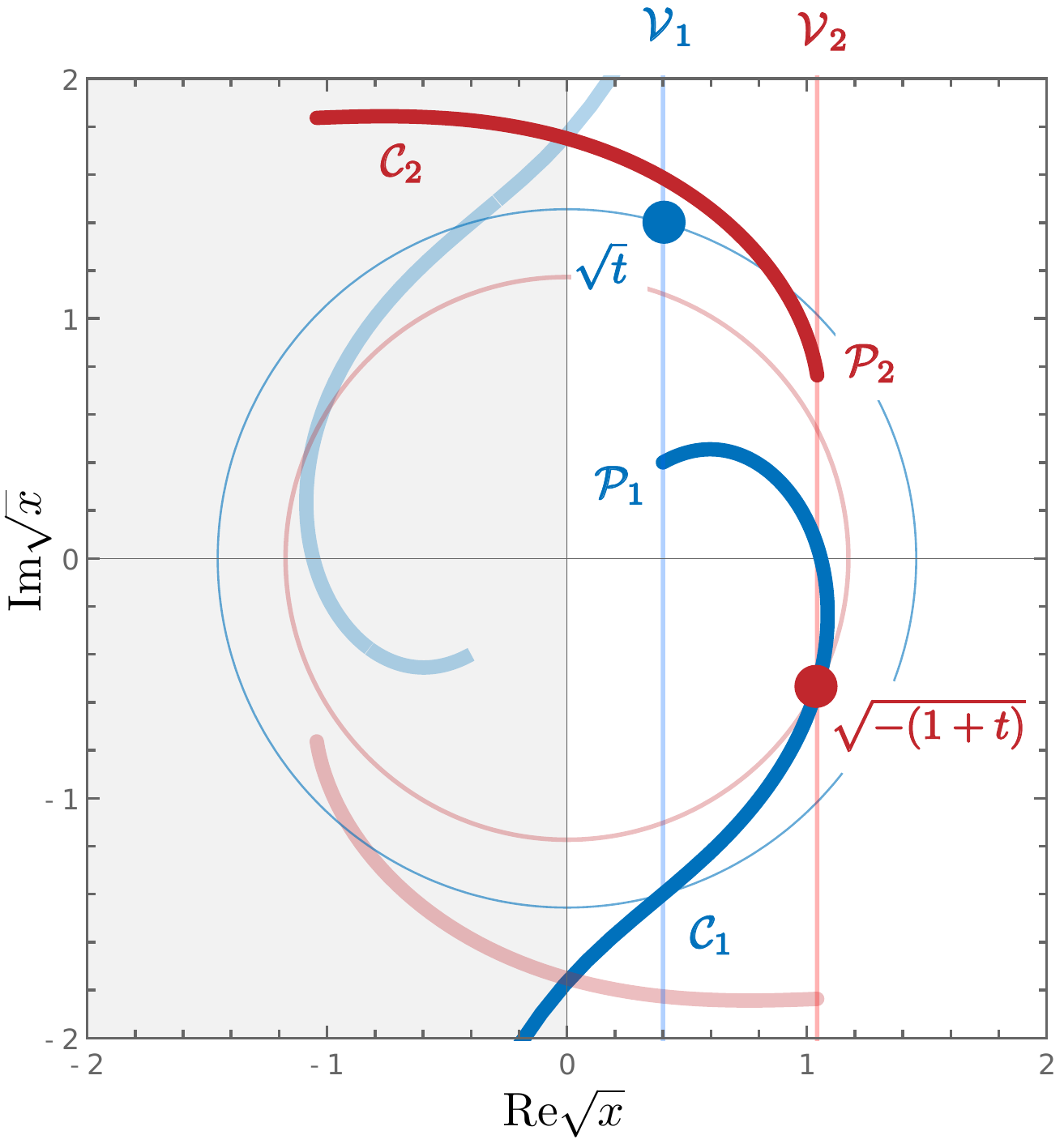}}
        \caption{Exemplary alignment of the two cuts $\mC_1$ and $\mC_2$ for $t=0.4+1.4i$ and $\alpha=1.3$,
        see text for a discussion.}
        \label{fig:generic-cut}
     \end{figure}

  \pagebreak

        \begin{appendix}

        \section{Contour deformation with two cuts} \label{app:cd}

        In this appendix we provide details on our contour deformation procedure in the calculation
        of the scattering amplitude. As discussed in Sec.~\ref{sec:half-offshell-amp},
        the complication in this case arises from the appearance of the two cuts
        \begin{equation}
        \begin{split}
            \mathcal{C}_1 &= f_\pm(t,1,z)\,, \\
            \mathcal{C}_2 &= f_\pm(-(1+t),\beta,\sqrt{1-z^2}\,y)
        \end{split}
        \end{equation}
        in the complex $\sqrt{x}$ plane which must be circumvented,
        where the function $f_\pm$ is given by
            \begin{equation}\label{f-cut-1}
               f_\pm(t,\alpha,z) = \sqrt{t}\,z \pm i\sqrt{t(1-z^2)+\alpha^2}\,.
            \end{equation}

     \begin{figure*}[t]
     \center{
     \includegraphics[width=0.95\textwidth]{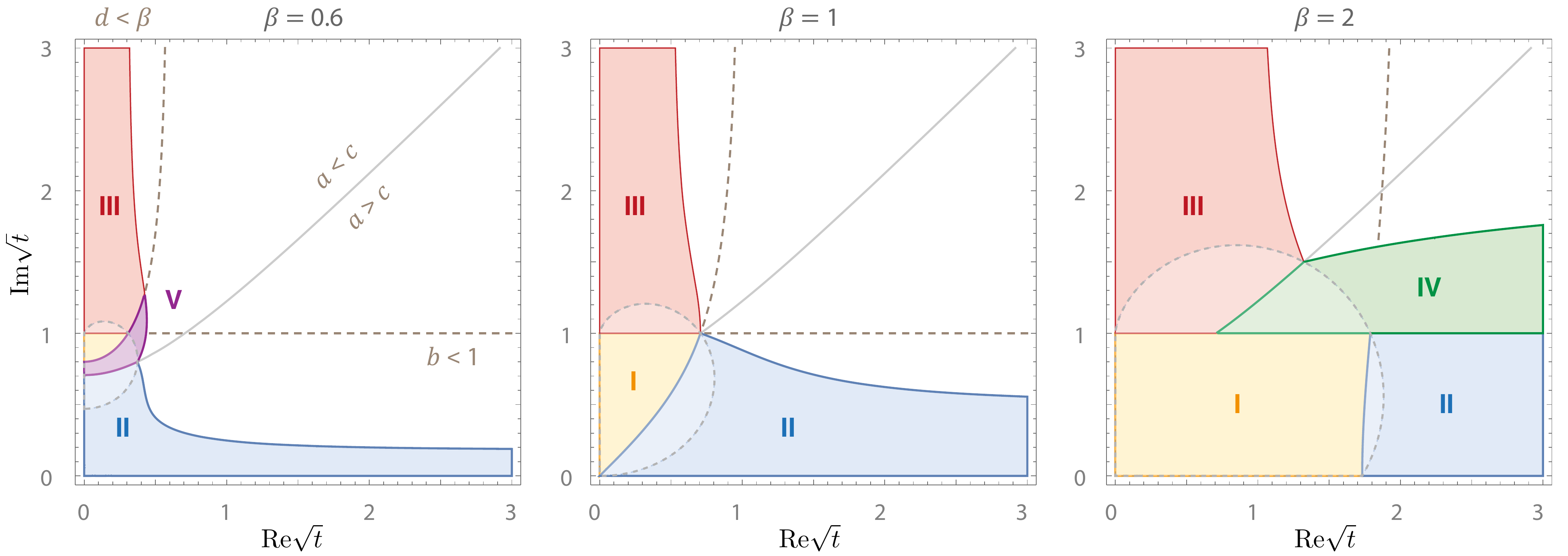}}
        \caption{Regions in the complex $\sqrt{t}$ plane which require a different contour deformation.}
        \label{fig:sqrt-t-regions}
     \end{figure*}

        With $\sqrt{t}$ in the upper right quadrant,
        $\text{Im}\,[-(1+t)]<0$ and therefore $\sqrt{-(1+t)}$ is in the lower right quadrant.
        Hence we can write
        \begin{equation}
           \sqrt{t} = a + ib, \qquad
           \sqrt{-(1+t)} = c - id
        \end{equation}
        where $a, b, c, d >0$ and
        \begin{align}
           c &= \sqrt{\frac{A-B}{2}}\,, \qquad
           d = \sqrt{\frac{A+B}{2}}\,,  \nonumber \\
           A^2 &= (1+\text{Re}\,t)^2 + (\text{Im}\,t)^2 = (1+a^2-b^2)^2 + 4a^2 b^2\,, \nonumber \\
           B &= 1+\text{Re}\,t = 1+a^2-b^2
        \end{align}
        with $A>0$, $|B|<A$.

        The typical situation in the complex $\sqrt{x}$ plane is illustrated in Fig.~\ref{fig:generic-cut}.
        The point $\sqrt{t}$ in the upper right quadrant defines the vertical line $\mV_1$ and a circle
        with radius $R_1 = \sqrt{a^2+b^2}$.
        The point $\sqrt{-(1+t)}$ defines the line $\mV_2$ and a circle
        with radius
        $R_2 = \sqrt{c^2+d^2} = \sqrt{A}$.
        The two relevant cuts are then given by
        \begin{equation}\label{c1}
        \begin{split}
           \mC_1 &= f_-((a+ib)^2,1,z)\,,  \\
           \mC_2 &= f_+((c-id)^2,\beta,z) 
        \end{split}
        \end{equation}
        where $z \in (-1,1)$ is a generic variable parametrizing the cuts and
        the opposite branches are not relevant for the further discussion.
        At $z=1$, the cuts start on $\mV_1$ and $\mV_2$ at the respective points
        \begin{equation}
           \mP_1  = a+i(b-1)\,, \quad
           \mP_2  = c + i(\beta-d)\,.
        \end{equation}
        As $z$ decreases, $|\mC_1|$ and $|\mC_2|$ increase.
        At $z=0$, $\mC_1$ passes through $\sqrt{-(1+t)}$ at $\mV_2$.
        At the point $\sqrt{t^\ast}$, again at $\mV_1$, the cut $\mC_1$ leaves the circle with radius $R_1$
        and continues until it reaches its endpoint $-a-i(b+1)$.
        The cut $\mC_2$ in Fig.~\ref{fig:generic-cut} already starts outside of its circle with radius $R_2$
        and progresses until its endpoint $-c + i(\beta+d)$.

        A proper integration path $\sqrt{x}=\mI$ in Fig.~\ref{fig:generic-cut} would start at the origin, pass through the region between $\mP_1$ and $\mP_2$
        and return to the real axis afterwards. In addition, it must satisfy the constraints
        arising from the exchange-particle cut $\mC_3$,
        namely that $\text{Re}\,\mI$ and $|\mI|$ must never decrease along the path.
        For example, once $\mI$ has picked up a nonvanishing real part
        it is no longer possible to return to the imaginary axis.

        Depending on the alignment of the cuts, different path deformations may be necessary.
        To this end we identify the following regions:
        \begin{itemize}
        \item $\mC_1$ never crosses the positive real axis if $|\text{Im}\sqrt{t}|<1$, cf.~Eq.~\eqref{cond-0},
              which entails $b<1$.

        \item $\mC_2$ never crosses the real axis if $|\text{Im}\sqrt{-(1+t)}|<\beta$, which entails
              \begin{equation}
                 d<\beta \quad \Leftrightarrow \quad b> \beta\,\sqrt{\frac{a^2-(\beta^2-1)}{\beta^2-a^2}}\,.
              \end{equation}

        \item The line $\mV_1$ is on the left of $\mV_2$ if
              \begin{equation}
                 |t|>|1+t| \quad\Leftrightarrow\quad \text{Re}\,t<-\tfrac{1}{2}\,,
              \end{equation}
              which corresponds to
              \begin{equation}
                 a<c \quad \Leftrightarrow \quad b > \sqrt{\tfrac{1}{2} + a^2}
              \end{equation}
              and implies $R_1>R_2$. If $\mV_1$ is on the right, the situation is reversed and $a>c$.

        \item If $a<c$, a contour deformation is possible as long as
              $\mP_2$ does not touch $\mC_1$, which leads to the condition
              \begin{equation}\label{cond-a<c}
                  \beta > \frac{G-\sqrt{G^2-8a^2\,(c^2+d^2)}}{2a}
              \end{equation}
              with $G=bc+3ad$.

        \item If $a>c$, a contour deformation is possible as long as
              $\mP_1$ does not touch $\mC_2$, which corresponds to
              \begin{equation}\label{cond-a>c}
                  \beta > \sqrt{\frac{2c\left(a^2+(1-b)^2\right)}{c(1-b)+ad}}\,.
              \end{equation}

        \end{itemize}

     \begin{figure*}[t]
     \center{
     \includegraphics[width=0.97\textwidth]{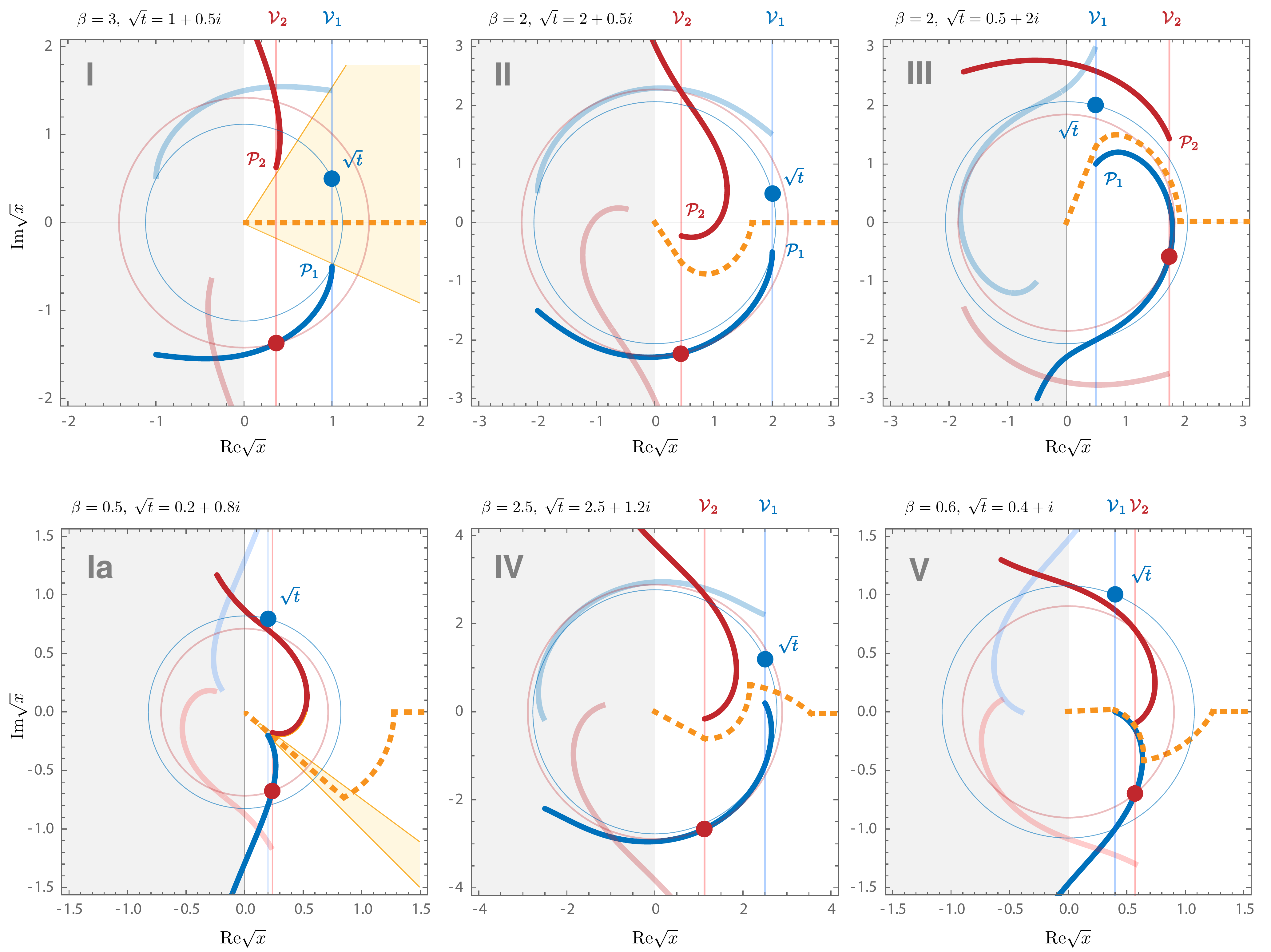}}
        \caption{Exemplary cut alignments in the complex $\sqrt{x}$ plane for the different regions,
                 together with the corresponding integration paths (dashed lines).}
        \label{fig:cuts-examples-2}
     \end{figure*}

        Based on these constraints, we divide up the complex $\sqrt{t}$ plane
        into regions which are distinguished by a different contour deformation procedure.
        They are plotted in Fig.~\ref{fig:sqrt-t-regions} for three values of the mass ratio $\beta$,
        where the constraints $b=1$ and $d=\beta$ correspond to the dashed lines and $a=c$ to the solid line.
        Fig.~\ref{fig:cuts-examples-2} illustrates exemplary configurations for each region:

        \medskip

        \textbf{Region I} ($b<1$, $d<\beta$): No contour deformation is necessary because
        none of the cuts cross the real axis, as exemplified in the upper left panel of Fig.~\ref{fig:cuts-examples-2}.

        \medskip

        \textbf{Region Ia}: The simplest extension is to test whether the cuts $\mC_1$ and $\mC_2$
        can be passed by a straight line starting at the origin, which is true if $\arg\mP_2 > \arg\mP_1$.
        Because both points are on the right half plane, we can drop the $\arctan$ and the condition becomes
        \begin{equation}
           a\,(\beta-d)+c(1-b)>0\,.
        \end{equation}
        An exemplary configuration is plotted in the bottom left panel of Fig.~\ref{fig:cuts-examples-2}.
        The optimal integration path is the average of the two lines
        connecting the origin with $\mP_1$ and $\mP_2$, respectively, where we
        integrate up to a radius which is larger than both $R_1$ and $R_2$
        before going back to the real axis.
        The resulting region in the complex $\sqrt{t}$ plane is shown by the dashed opaque areas in Fig.~\ref{fig:sqrt-t-regions}.
        They enclose region~I but also extend into the other regions discussed below.

        \medskip

        \textbf{Region II} ($b<1$, $d>\beta$, $a>c$): Only $\mC_2$ crosses the real axis,
        as illustrated in the top center panel of Fig.~\ref{fig:cuts-examples-2}. Here the path deformation is still simple
        because we only need to circumvent $\mC_2$ and can return to the real axis afterwards.

        To find the optimal integration path, consider the function $f_+((c-id)^2,\alpha,z)$.
        For $\alpha=\beta$ this is just the cut $\mC_2$.
        As we decrease $\alpha$, we deform its shape; $\mP_2$ slides down along the line $\mV_2$
        until the contour eventually touches the cut $\mC_1$ at the point $\mP_1$.
        The corresponding value of $\alpha$ is given by the r.h.s. of Eq.~\eqref{cond-a>c}.
        Decreasing $\alpha$ further, the contour crosses the cut $\mC_1$ and for
        $\alpha = 0$ it becomes the half-circle starting at $c-id$.
        Therefore, the choice
        \begin{equation}
        \begin{split}
          \mI_+(z) &= f_+((c-id)^2,\alpha_+,z)\,, \\
          \alpha_+ &= \sqrt{\frac{1}{2}\left[ \beta^2 + \frac{2c\left(a^2+(1-b)^2\right)}{c(1-b)+ad}\right]}\,,
        \end{split}
        \end{equation}
        which averages over the cut $\mC_2$ and the contour touching $\mP_1$,
        is guaranteed to lie between the cuts $\mC_1$ and $\mC_2$.

        Our integration path thus proceeds from the origin to the point
        \begin{equation}
           \mI_+(z=1)=c+i(\alpha_+-d)
        \end{equation}
        on $\mV_2$,
        goes along $\mI_+(z)$ until it reaches the real axis at
        \begin{equation}
          \mI_+(z_+)= \sqrt{R_2^2-\alpha_+^2}\,, \quad
          z_+ = \frac{c}{\sqrt{R_2^2-\alpha_+^2}}
        \end{equation}
        and from there continues to $\sqrt{x} \to \infty$.
        Because the real part and modulus of the contour always increase along that path,
        the constraints from the exchange-particle cut $\mC_3$ are satisfied.
        In this way we can cover the entire region~II without crossing any branch cut.

        \medskip

        \textbf{Region III} ($b>1$, $d<\beta$, $a<c$): This is the opposite case where only $\mC_1$ crosses the real axis
        (Fig.~\ref{fig:generic-cut} and top right panel in Fig.~\ref{fig:cuts-examples-2}).
        Here we employ the function $f_-((a+ib)^2,\alpha,z)$.
        For $\alpha=1$ this is the cut $\mC_1$ and for $\alpha=0$ it becomes the circle passing through $a+ib$.
        As we decrease $\alpha$, $\mP_1$ will slide up along the line $\mV_1$ until
        the contour touches the point $\mP_2$ of the cut $\mC_2$.
        The resulting value for $\alpha$, however, is not the one in Eq.~\eqref{cond-a<c}, which was
        obtained by equating $\mP_2$ and $\mC_1$ with $\alpha=1$.
        Instead, we need to keep $\alpha$ general which results in
        \begin{equation}
        \begin{split}
           \mI_-(z) &= f_-((a+ib)^2,\alpha_-,z)\,, \\
           \alpha_- &= \sqrt{1+\frac{\bar\beta}{2}\,\frac{a\,(2R_2^2+\bar\beta^2)-\bar\beta\,G}{a(\bar\beta-d)+bc}} \,,
        \end{split}
        \end{equation}
        where $G=bc+3ad$ and
        \begin{equation}
           \bar\beta = \min \left( \beta, d+\sqrt{a^2+b^2-c^2} \right).
        \end{equation}
         Here we took care of another subtlety:
        If $|\mP_2|>R_1$ (which is the case in Fig.~\ref{fig:cuts-examples-2}),
        then the cut $\mC_2$ no longer reaches $f_-((a+ib)^2,\alpha,z)$
        which remains within the circle with radius $R_1$.
        In that case we replace $\mP_2$ by the intersection of that circle with $\mV_2$,
        which from $|\mP_2|=R_1$ corresponds to $\beta = d+\sqrt{a^2+b^2-c^2}$
        and leads to the above definition. The analogous situation for region II can never happen because $|\mP_1| < R_2$.

        The integration path then proceeds from the origin to
        \begin{equation}
           \mI_-(z=1)=a+i(b-\alpha_-)\,,
        \end{equation}
        on $\mV_1$, goes along $\mI_-(z)$ until it reaches the real axis at
        \begin{equation}
          \mI_-(z_-)= \sqrt{R_1^2-\alpha_-^2}\,, \quad
          z_- = \frac{a}{\sqrt{R_1^2-\alpha_-^2}}
        \end{equation}
        and from there continues to $\sqrt{x} \to \infty$.
        The constraints from the exchange-particle cut $\mC_3$ are again satisfied, so
        we can cover the entire region III without crossing any branch cut.

        \medskip

        \textbf{Region IV} ($b>1$, $a>c$): This case is more complicated and illustrated in the bottom center of Fig.~\ref{fig:cuts-examples-2}.
        Here $\mC_1$ crosses the real axis; whether $\mC_2$ also crosses or not is not relevant.
        The integration path is similar as in Region II, except that
        one must still circumvent the cut $\mC_1$ after passing the real axis.
        Because the real part of the contour must never decrease, one can stay on $\mI_+(z)$ only as long as it does not become vertical.
        At that point we switch to another contour that brings us back to the real axis.
        In Fig.~\ref{fig:sqrt-t-regions} one can see that Region IV only appears  for $\beta>1$.

     \begin{figure}[t!]
     \center{
     \includegraphics[width=0.9\columnwidth]{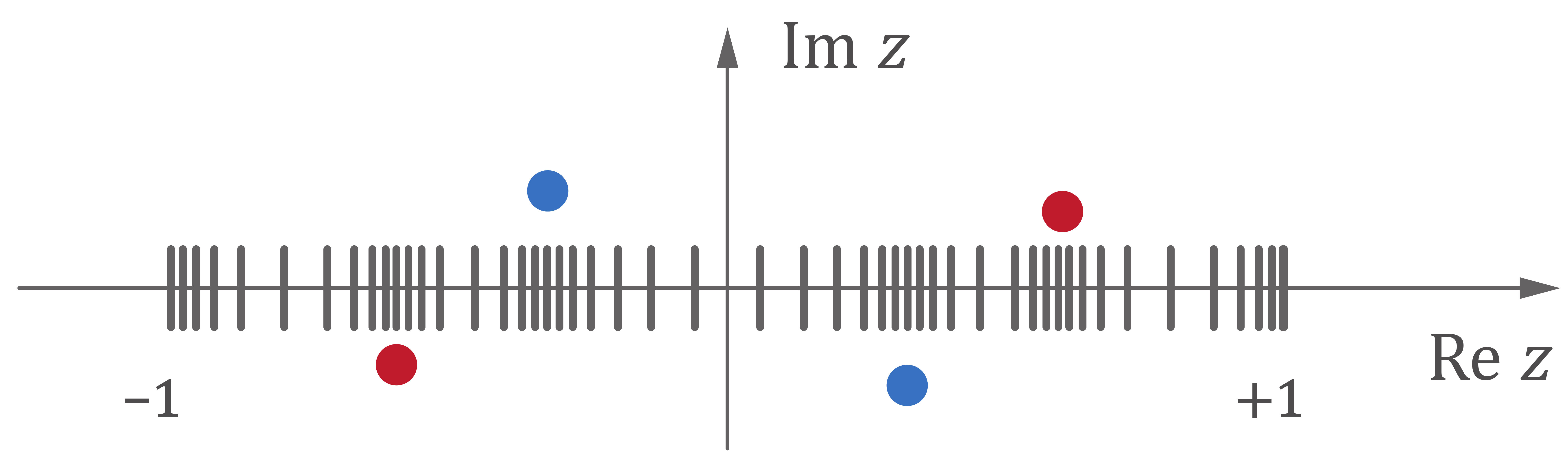}}
        \caption{Adaptive grid in the variable $z\in (-1,1)$. The points show exemplary pole locations from the integrand
                 in the complex $z$ plane.}
        \label{fig:z-integrand}
     \end{figure}

        \pagebreak

        \textbf{Region V} ($d>\beta$, $a<c$) is the opposite case shown in the bottom right of Fig.~\ref{fig:cuts-examples-2}.
        Here $\mC_2$ crosses the real axis, whereas $\mC_1$ may or may not intersect with the real axis.
        The integration path is similar to Region III;
        once again, one must switch contours if $\mI_-(z)$ becomes vertical.
        One can see in Fig.~\ref{fig:sqrt-t-regions} that Region V only covers a small portion of the complex $\sqrt{t}$ plane
        and it also only appears  for $\beta<1$.

        \medskip

        The union of all Regions I--V covers the entire accessible area in the complex $\sqrt{t}$ plane defined by Eqs.~(\ref{cond-a<c}--\ref{cond-a>c}).
        Region Ia also extends into the other areas and
        provides a useful cross check since in the overlap regions one can test different
        contour deformation strategies.

        In principle there are many possible ways to generalize or even automatize the contour deformation.
        For example, outside of Region Ia one could use the two rays connecting the origin with
        $\mP_1$ and $\mP_2$ to define inner and outer zones and construct integration paths accordingly.
        Alternatively, one could connect the tangent on one cut with the endpoint of the other
        and then integrate along straight lines. Such techniques
        can be useful in more general situations, for example when the ingredients
        of the equation are nonperturbative and contain not only tree-level poles but also poles or cuts in the complex plane.

        Finally, what turned out very useful was a pole analysis in the complex $z$ plane,
        which is the angular integration variable in the BSE~\eqref{bse-3} or scattering equation~\eqref{sc-eq-3}.
        The $z$ integration goes from $z=-1$ to $z=+1$ and for a given path deformation in $\sqrt{x}$
        the poles of the integrands never cross the integration path in $z$. Still, the closer the path in $\sqrt{x}$
        comes to one of those singularities (e.g. in region Ia of Fig.~\ref{fig:cuts-examples-2}, when it passes through the cuts),
        the closer the poles in the complex $z$ plane come to the real axis and therefore
        the integrand as a function of $z$ varies strongly in the vicinity of those poles.
        Here it is useful to perform an adaptive integration by splitting the $z$ integration into intervals
        and accumulate the grid points around the nearest singularities as sketched in Fig.~\ref{fig:z-integrand}.
        For example, for the propagator poles corresponding to $\mC_1$ their
        locations are determined by solving $\sqrt{x}=f_\pm(t,1,z)$ for $z$.
        In that way we typically gain a factor $\sim 10^2 \dots 10^3$ in CPU time while maintaining the same numerical accuracy.

    \end{appendix}

\bibliographystyle{apsrev4-1-mod}

\bibliography{lit-resonances}

\end{document}